\documentclass[12pt,a4paper]{article}
\usepackage[textwidth=15cm]{geometry}
\usepackage{authblk}
\usepackage{amsmath} 
\usepackage{bm}        
\usepackage{amssymb}   
\usepackage[small]{caption}
\usepackage{epsf,epsfig,graphicx,color,hyperref}

\newcommand{\vevr}{\left< r \right>}
\newcommand{\vevh}{\left< \varphi \right>}

\newcommand{\vM}{v_{\rm M}}
\newcommand{\al}{\alpha}
\newcommand{\Gev}{\rm GeV}
\newcommand{\Mev}{\rm MeV}

\newcommand{\Tev}{\rm TeV}

\newcommand{\vk}{\vec{k}}

\newcommand{\vl}{\vec{l}}

\newcommand{\vsigma}{\vec{\sigma}}
\newcommand{\tilom}{\tilde{\omega}}
\newcommand{\twopi}{\left( 2 \pi \right)}
\newcommand{\mathM}{\mathcal{M}}
\newcommand{\mathA}{\mathcal{A}}

\begin{document}

\title{Supernovae and Weinberg's Higgs Portal Dark Radiation and Dark Matter}

\author{Huitzu Tu\thanks{huitzu2@gate.sinica.edu.tw}\,}
\author{Kin-Wang Ng\thanks{nkw@phys.sinica.edu.tw}}
\affil{Institute of Physics, Academia Sinica, Taipei 11529, Taiwan}

\maketitle

\begin{abstract}

The observed burst duration and energies of the neutrinos from Supernova 1987A 
strongly limit the possibility of any weakly-interacting light particle
species being produced in the proto-neutron star (PNS) core and leading to efficient 
energy loss. 
We reexamine this constraint on Weinberg's Higgs portal model, in which the dark
radiation particles (the Goldstone bosons) and the dark matter candidate 
(a Majorana fermion) interact with Standard Model (SM) fields solely through the   
mixing of the SM Higgs boson and a light Higgs boson.
In order for the Goldstone bosons to freely stream out of the PNS core region, the
Higgs portal coupling has to be about a factor of $4$--$9$ smaller than the current 
collider bound inferred from the SM Higgs invisible decay width.
We find that in the energy loss rate calculations, results obtained by 
using the one-pion exchange (OPE) approximation and the SP07 global fits for the 
nucleon-nucleon total elastic cross section differ only by a factor $\lesssim 3$.
The SN 1987A constraints surpass those set by laboratory experiments or by the 
energy loss arguments in other astrophysical objects such as the gamma-ray bursts,
even with other nuclear uncertainties taken into account.
Furthermore, the SN 1987A constraints are comparable to bounds from the latest 
dark matter direct search for low-mass WIMPs ($\lesssim 10~\Gev$.)

\end{abstract}

\maketitle

\section{Introduction}
\label{sec:intro}

SN 1987A was a type II supernova discovered on February 24, 1987 by Shelton, Duhalde
and Jones.
The progenitor star was Sanduleak $-69^\circ$ 202, a blue supergiant in the 
Large Magellanic Cloud.
Thanks to its proximity of about $51~{\rm kpc}$ to the Earth, neutrino burst events
from the core collapse of the progenitor star could be recorded at the underground
laboratories Irvine-Michigan-Brookhaven (IMB), Kamiokande II, and Baksan 
separately~\cite{Raffelt:1996wa}.
The observed burst duration of about $12$ seconds, individual energies up to $40~\Mev$, 
as well as the integrated total energy of $\mathcal{O} (10^{53}~{\rm erg})$, 
confirmed the standard picture of neutrino cooling of the proto-neutron 
star (PNS)~\cite{Fischer:2009af,Mueller:2014rna,Camelio:2017nka}.
A proto-neutron star is formed when the collapsing 
stellar core of the progenitor star reaches nuclear saturation density.
Being initially hot and lepton rich, the PNS keeps contracting as it cools and 
deleptonise, to become a neutron star as the final supernova remnant. 
See Refs.~\cite{Prakash:1996xs,Pons:1998mm,Nicotra:2005fj} for the PNS structure 
and the evolution, and Ref.~\cite{Janka:2017vlw} for the most recent review on 
neutrino emission from supernovae.

Emission of light exotic particles in nuclear interactions in the PNS core
have been considered exhaustively in the literature, notably the 
axions~\cite{Raffelt:1987yt,Turner:1987by,Mayle:1987as,Brinkmann:1988vi,Janka:1995ir}, 
right-handed neutrinos~\cite{Raffelt:1987yt}, 
Kaluza-Klein gravitons~\cite{Hanhart:2000er,Hanhart:2001fx,Hannestad:2003yd}, 
Kaluza-Klein dilatons~~\cite{Hanhart:2000er}, 
unparticles~\cite{Hannestad:2007ys,Freitas:2007ip}, 
dark photons~\cite{Chang:2016ntp}, dark matter~\cite{Guha:2015kka}, 
dilation~\cite{Ishizuka:1989ts}, saxion~\cite{Arndt:2002yg} etc.
Simulations of PNS in the neutrino-emitting phase were done in 
Refs.~\cite{Keil:1996ju,Fischer:2016cyd} for the axion, 
and in Ref.~\cite{Hanhart:2001fx} for the KK-gravitons.
By comparing the predicted neutrino burst signals with the SN 1987A observations,
very stringent constraints were obtained on the properties of the exotic particles.
For a quick comparison without invoking simulations, Raffelt has derived a bound
on the emissivity of light exotic particles based on the argument that they should
not affect the total cooling time significantly~\cite{Raffelt:1990yz,Raffelt:2006cw}. 

In this work we shall reexamine the SN 1987A constraints on Weinberg's Higgs 
portal model~\cite{Weinberg:2013kea}, which was proposed to account for the dark radiation in the early universe. 
The effect of the dark radiation on the cosmic microwave background (CMB) data
is parametrised as the contribution to the effective number of light neutrino species
$N_{\rm eff}$. 
The conflict between the value of the Hubble constant $H_0$ from the Planck CMB data 
and local determination may be remedied by assuming an addition of 
$\Delta N_{\rm eff} = 0.4$--$1$ to the standard value of $N_\nu = 3.046$
by the dark radiation component~\cite{Riess:2016jrr}
(see, however, also Ref.~\cite{Heavens:2017hkr}.)
In this model, Weinberg considered a global $U (1)$ continuous symmetry associated 
with the conservation of some quantum number, and introduced a complex scalar field 
to break it spontaneously.
The radial field of the complex scalar field acquires a vacuum expectation value
(vev), and mixes with the Standard Model (SM) Higgs field.
The Goldstone bosons arising from the symmetry breaking would be massless, and their characteristic derivative coupling would make them very weakly-interacting 
at sufficiently low temperatures.
The latter property is crucial, since the Goldstone bosons must decouple from the early
universe thermal bath at the right moment so that their temperature is a fraction of 
that of the neutrinos (see e.g. Ref.~\cite{Ng:2014iqa}.)
Collider phenomenology of Weinberg's Higgs portal model has been investigated in 
Refs.~\cite{Cheung:2013oya,Anchordoqui:2013bfa}.
Weinberg has also extended this minimal set-up to include a Majorana fermion 
as a Weakly-Interacting Massive Particle (WIMP) dark matter candidate.
Ref.~\cite{Anchordoqui:2013bfa} has shown that results of the dark matter direct 
search experiments LUX~\cite{Akerib:2016vxi} provide very strong constraints, 
which are slightly strengthened by the XENON1T experiment~\cite{Aprile:2017iyp} very
recently.

Previously we have examined energy losses due to the emission of Weinberg's 
Goldstone bosons in a post-collapse supernova core~\cite{Keung:2013mfa} in the limit
of large radial field mass.
Subsequently we scrutinised the production and propagation of Weinberg's Goldstone 
bosons in the initial fireballs of gamma-ray bursts for more general 
cases~\cite{Tu:2015lwv}.
In this work we extend our previous analysis and consider in greater detail 
Goldstone boson production by nuclear bremsstrahlung processes in the proto-neutron 
star core of SN 1987A.
In Sec.~\ref{sec:model} we briefly review Weinberg's Higgs portal model for 
dark radiation and dark matter.
In Sec.~\ref{sec:GBproduction} we calculate energy loss rate due to Goldstone 
boson emission by two methods, i.e. using the one-pion exchange approximation and using experimental data of low-energy nucleon collisions.
In Sec.~\ref{sec:GBmfp} we estimate the mean free path of the Goldstone bosons
as a function of their emission energies, and determine the free-streaming 
requirements.
Our results in these two sections are then used in Sec.~\ref{sec:constraints} 
to derive supernova constraints on Weinberg's Higgs portal model by invoking 
Raffelt's criterion. 
We then confront our SN 1987A constraints with those from accelerator experiments,
gamma-ray burst observations, and dark matter direct search experiments.
In Sec.~\ref{sec:summary} we summarise our work.

\section{Weinberg's Higgs portal model}
\label{sec:model}

In this section we briefly summarise Weinberg's model~\cite{Weinberg:2013kea} 
following the convention of Refs.~\cite{Cheung:2013oya,Keung:2013mfa}.
Consider the simplest possible broken continuous symmetry, a global $U (1)$ symmetry
associated with the conservation of some quantum number $W$.
A single complex scalar field $S (x)$ is introduced for breaking this symmetry
spontaneously. 
With this field added to the Standard Model (SM), the Lagrangian is 
\begin{equation}
\label{eq:Lagrangian1}
   {\mathcal L} = \left(\partial_\mu S^\dagger \right) \left(\partial^\mu S \right)
   + \mu^2 S^\dagger S - \lambda (S^\dagger S)^2 - g (S^\dagger S) 
   (\Phi^\dagger \Phi) + {\mathcal L}_{\rm SM}\, .
\end{equation}
where $\Phi$ is the SM Higgs doublet, $\mu^2$, $g$, and $\lambda$ are real constants,
and $\mathcal{L}_{\rm SM}$ is the usual SM Lagrangian.
One separates a massless Goldstone boson field $\alpha (x)$ and a massive radial field
$r (x)$ in $S (x)$ by defining
\begin{equation}
   S (x) = \frac{1}{\sqrt{2}} \left(\vevr + r (x) \right)\, e^{2 i \al (x)}\, .
\end{equation}
where the fields $\alpha (x)$ and $r (x)$ are real.
In the unitary gauge, one sets $\Phi^{\rm T} = \left(0, \vevh + \varphi (x) \right) 
/\sqrt{2}$ where $\varphi (x)$ is the physical Higgs field.
The Lagrangian in Eq.~(\ref{eq:Lagrangian1}) thus becomes
\begin{eqnarray}
   \mathcal{L} &=& \frac{1}{2} \left(\partial_\mu r \right) \left(\partial^\mu r \right)
   + \frac{1}{2} \frac{\left(\vevr + r \right)^2}{\vevr^2} 
   \left(\partial_\mu \al \right) \left(\partial^\mu \al \right) +
   \frac{\mu^2}{2} \left(\vevr + r \right)^2 \nonumber \\
   && - \frac{\lambda}{4} \left(\vevr + r \right)^4 - \frac{g}{4}
   \left(\vevr + r \right)^2 \left(\vevh + \varphi \right)^2 + \mathcal{L}_{\rm SM}\, ,
\end{eqnarray}
where the replacement $\al (x) \rightarrow \al (x) / \left( 2 \vevr \right)$ was made 
in order to achieve a canonical kinetic term for the $\al (x)$ field.
The two fields $\varphi$ and $r$ mix due to the $g (S^\dagger S) (\Phi^\dagger \Phi)$
term, with their mixing angle given by
\begin{equation}
\label{eq:mixingangle}
   \tan 2 \theta = \frac{2 g \vevh \vevr}{m^2_H - m^2_h}\, ,
\end{equation}
where $m_H$ and $m_h$ are the masses of the two resulting physical Higgs bosons 
$H$ and $h$, respectively.
The heavier one is identified with the SM Higgs boson with $m_H = 125~\Gev$, while
the lighter one is assumed to have a mass in the range of MeV to hundreds of MeV.
In this model, the interaction of the Goldstone bosons with the SM fields arises 
entirely through the SM Higgs boson in the mixing of the $\varphi$ and $r$ fields. 
The light Higgs boson $h$ decays dominantly to a pair of Goldstone bosons, with
the decay width given by
\begin{equation}
   \Gamma_h = \frac{1}{32 \pi} \frac{m^3_h}{\vevr^2}\, .
\end{equation}
When kinematically allowed, there is also a probability for $h$ decaying into a 
pair of SM fermions as well as a pair of pions~\cite{Tu:2015lwv}.

The Higgs effective coupling to nucleons, $f_N m_N /\vevh \equiv g_{N N H}$, has been
calculated for the purpose of investigating the sensitivities of the 
dark matter direct detection experiments~\cite{Drees:1993bu,Jungman:1995df,Hisano:2011cs,Cheng:2012qr,Cline:2013gha}.
Ref.~\cite{Cheng:2012qr} found $g_{N N H} = 0.0011$, 
which corresponds to $f_N \simeq 0.288$.
It was pointed out in Ref.~\cite{He:2013suk} that the effective Higgs-nucleon
coupling has a wide range of values, $0.0011 \leq g_{N N H} \leq 0.0032$, 
due to uncertainties in the pion-nucleon sigma term.
The authors of Ref.~\cite{Cline:2013gha} have done a statistical analysis to infer 
the value of $f_N$ from more up-to-date lattice evaluations of the nucleon 
matrix elements.
By exploiting two possible statistical distributions for the strangeness matrix 
element, they found $f_N = 0.3 \pm 0.03$ and $f_N = 0.3 \pm 0.01$ at the $68\%$
confidence level, respectively.

This model is also extended to include a dark matter candidate by
adding one Dirac field
\begin{equation}
   \mathcal{L}_\psi = i \bar{\psi} \gamma \cdot \partial \psi - m_\psi \bar{\psi} \psi 
   - \frac{f_\chi}{\sqrt{2}} \bar{\psi}^c \psi S^\dagger - 
   \frac{f^\ast_\chi}{\sqrt{2}} \bar{\psi} \psi^c S\, , 
\end{equation}
and assigning a charge $U (1)_W = 1$ for it.
One expresses the field as $\psi (x) = \psi^\prime (x) e^{i \alpha (x)}$, and
expands the Lagrangian after the radial field achieves a vev (for details see
Ref.~\cite{Anchordoqui:2013bfa}.)
Diagonalising the $\psi^\prime$ mass matrix generates the mass eigenvalues
\begin{equation}
   m_\pm = m_\psi \pm f_\chi \vevr\, ,
\end{equation}
for the two mass eigenstates $\psi_\pm$, which are Majorana fermions.
The Lagrangian is now
\begin{eqnarray}
   \mathcal{L}_\psi &=& \frac{i}{2} \bar{\psi}_{\pm} \gamma \cdot \partial \psi_{\pm}
   - \frac{1}{2} m_\pm \bar{\psi}_\pm \psi_\pm 
   - \frac{i}{4 \vevr} \left(\bar{\psi}_+ \gamma \psi_- 
   - \psi{\psi}_- \gamma \psi_+ \right) \cdot \partial \al \nonumber \\
   && - \frac{f_\chi}{2}\, r\, \left(\bar{\psi}_+ \psi_+ -
   \bar{\psi}_- \psi_- \right)\, ,
\end{eqnarray}
and one needs to use the massive representation 
$r = \cos \theta\, h + \sin \theta\, H$ for the interactions of $\psi_\pm$. 
The heavier fermion decays into the lighter fermion by emitting a Goldstone boson, 
while the lighter one is stable due to unbroken reflection symmetry.
The latter can thus play the role of the WIMP dark matter, 
with mass $m_- \equiv m_\chi$ in the range of GeV to TeV.
Its relic density has been calculated in Ref.~\cite{Anchordoqui:2013pta}. 

Model parameters in the minimal set-up are $m_h$, $g$, and $\vevr$, and
including $m_\chi$ and $f_\chi$ in the extended version.
From the SM Higgs invisible decay width, a collider bound on the Higgs portal 
coupling  
\begin{equation}
\label{eq:colliderbound}
   g < 0.011\, ,
\end{equation}
has been derived in Ref.~\cite{Cheung:2013oya}.
In the future, the International Linear Collider (ILC) may reach a sensitivity of 
constraining the branching ratio of SM Higgs invisible decays to 
$< 0.4$--$0.9\%$~\cite{Bechtle:2014ewa} in the best scenarios.
If this can be realised, the collider bound on the Goldstone boson coupling will be 
improved by a factor of $5 \sim 7$.
Experimental limits on meson invisible decay widths have also been turned into constraints on the $\varphi$-$r$ mixing angle in Ref.~\cite{Anchordoqui:2013bfa}, 
which we list in Sec.~\ref{sec:constraints}.
There is also the perturbativity condition, which requires for the quartic 
self-coupling of the $S$ field
\begin{equation}
\label{eq:perturbativity}
   \lambda = \frac{m^2_h}{\vevr^2} \leq 4 \pi\, .
\end{equation}
In Weinberg's Higgs portal model including the dark matter candidate, exclusion
limits on the WIMP-nucleon elastic cross section set by the null results of the 
direct search experiments have been found to put very strong bounds on the mixing 
angle in Ref.~\cite{Anchordoqui:2013bfa}.

\section{Goldstone boson production in proto-neutron star core}
\label{sec:GBproduction}

In the PNS core, the dominant Goldstone boson production channel is the nuclear bremsstrahlung processes $N N \rightarrow N N \al \al$.
Low-energy nuclear interactions have been studied quite thoroughly by various
experiments, while theoretical calculation remains a difficult task.
Taketani, Nakamura and Sasaki~\cite{Taketani:1951} suggested to divide the 
nuclear forces into three regions: classical (long-range), 
a dynamical (intermediate range), and a phenomenological or core (short-range) region.
In the classical region, the one-pion exchange (OPE) dominates the 
longest range part of the potential. 
In the intermediate range the two-pion exchange (TPE) is most important, 
where heavier mesons may also become relevant.
In the short-range region, multi-pion exchange, heavy mesons, quark-gluon exchanges
are expected to be responsible. 
At present $N N$ potentials calculated using the chiral effective field theory 
to the fifth order (${\rm N^{4} LO}$)~\cite{Entem:2017gor} 
and the sixth order (${\rm N^{5} LO}$)~\cite{Entem:2015xwa} are available, 
which can reproduce the experimental data to outstanding precision.
See e.g. Refs.~\cite{Machleidt:2001rw,Naghdi:2007ek,Machleidt:2011zz,Machleidt:2016vlh} for reviews on nucleon-nucleon interactions.

As for nuclear bremsstrahlung processes, in Refs.~\cite{Bacca:2008yr,Bacca:2015tva} neutrino pair production in core-collapse supernovae was studied using chiral effective 
field theory to the fourth order (${\rm N^{3} LO}$).
It was found that shorter-range noncentral forces significantly reduce 
the neutrino rates compared to the one-pion exchange (OPE) 
approximation~\cite{Brinkmann:1988vi,Friman:1978zq,Hannestad:1997gc}, which was
typically used in supernova simulations or in deriving supernova bounds on exotic particles.
More recently, Ref.~\cite{Bartl:2016iok} goes beyond the OPE approach and 
uses $T$-matrix based formalism from Ref.~\cite{Bartl:2014hoa} in their supernova simulations.
The approach of using phase shift data to fix the on-shell $NN$ scattering amplitudes 
and making the soft-radiation approximation has already been taken in 
Ref.~\cite{Hanhart:2000ae} much earlier.
It was found therein that the resultant rates are roughly a factor of four below 
earlier estimates based on an OPE $NN$ amplitude.

In this section we make the same comparison in Weinberg's Higgs portal model.

\subsection{Energy loss rate using one-pion exchange approximation}

The OPE contribution to the nuclear forces takes care of the long-range interactions
and the tensor force.
From the Lagrangian describing the pion coupling to nucleons
$\mathcal{L}_{\pi^0 N N} = - g_{\pi^0}\, \bar{\psi}\, i \gamma^5\, \tau_3\, \psi\, 
\varphi^{(\pi^0)}$, where $N = n$, $p$, the potential is 
\begin{equation}
   V_{\rm OPE} (\vk) = - \left(\frac{f_\pi}{m_\pi} \right)^2\,
   \frac{\left(\vsigma_1 \cdot \vk \right) \left(\vsigma_2 \cdot \vk \right)}
   {|\vk|^2 + m^2_\pi}\, \left(\vec{\tau}_1 \cdot \vec{\tau}_2 \right)\, ,
\end{equation}
with $\vec{k}$ the momentum exchange, and $\vsigma_j$ and $\vec{\tau}_i$ the spin and isospin operators of the incoming nucleons, respectively.
The neutral pion-nucleon coupling constant is 
$g^2_{\pi^0} / 4 \pi = \left( 2 m_N f_\pi / m_\pi \right)^2 / 
\left( 4 \pi\right) \approx 14$~\cite{Limkaisang:2001yz,Babenko:2016idp},
with $f_\pi \approx 1$.
In the one-pion exchange (OPE) approximation (see e.g. Ref.~\cite{Brinkmann:1988vi}), 
there are four direct and four exchange diagrams, corresponding to the Goldstone boson 
pairs being emitted by any one of the nucleons.
Summing all diagrams and expanding in powers of $\left(T / m_N \right)$, 
the amplitude for the nuclear bremsstrahlung processes
$N (p_1)\, N (p_2) \rightarrow N (p_3)\, N (p_4)\, \al (q_1)\, \al (q_2)$ 
is~\cite{Keung:2013mfa} 
\begin{eqnarray} 
\label{eq:Mbremsstrahlung}
    \sum_{\rm spins} |\mathcal{M}^{\rm OPE}_{N N \rightarrow N N \al \al}|^2 
    &\approx& 64 \left(\frac{f_N\, g m_N}{m^2_H} \right)^2 
    \left(\frac{2 m_N f_\pi}{m_\pi} \right)^4 \frac{(q_1 \cdot q_2)^2}
    {(q^2 - m^2_h)^2 + m^2_h \Gamma^2_h} \nonumber \\
    && \hspace{-4.6cm} \cdot\, \frac{(-2 q^2)^2\, m^2_N}{(2 p \cdot q)^4}\, 
    \Big\{\frac{\vert \vk \vert^4}{(\vert \vk \vert^2 + m^2_\pi)^2} + 
    \frac{\vert \vl \vert^4}{(\vert \vl \vert^2 + m^2_\pi)^2} +
    \frac{\vert \vk \vert^2 \vert \vl \vert^2 - 2 |\vk \cdot \vl|^2}
    {(\vert \vk \vert^2 + m^2_\pi) (\vert \vl \vert^2 + m^2_\pi)} + ... \Big\}\, ,
\end{eqnarray}
where $q \equiv q_1 + q_2$, and $k \equiv p_2 - p_4$ and $l \equiv p_2 - p_3$ are the 
$4$-momenta of the exchanged pion in the direct and the exchange diagrams, 
respectively.
In addition, Goldstone boson pairs can be emitted from the exchanged pion due to 
an effective Higgs-pion coupling.
The amplitude for this process is
\begin{eqnarray} 
\label{eq:MOPE_pion}
    \sum_{\rm spins} |\mathcal{M}^{\rm OPE\, (pion)}_{N N \rightarrow N N \al \al}|^2 
    &\approx& 4 \left(\frac{g}{m^2_H} \right)^2 
    \left(\frac{2 m_N f_\pi}{m_\pi} \right)^4 \frac{(q_1 \cdot q_2)^2}
    {(q^2 - m^2_h)^2 + m^2_h \Gamma^2_h}\, \left(\frac{2}{9} \right)^2 \nonumber \\
    && \hspace{-4.6cm} \cdot\, \left(q^2 + \frac{11}{2} m^2_\pi \right)^2 
    \Big\{\frac{k^2_1 k^2_2}{(k^2_1 - m^2_\pi)^2\, (k^2_2 - m^2_\pi)^2} + 
    \frac{l^2_1 l^2_2}{(l^2_1 - m^2_\pi)^2\, (l^2_2 - m^2_\pi)^2} \nonumber \\
    && \hspace{-4.6cm} + \frac{(k_1 \cdot k_2) (l_1 \cdot l_2) + ...}
    {(k^2_1 - m^2_\pi) (k^2_2 - m^2_\pi) 
    (l^2_1 - m^2_\pi) (l^2_2 - m^2_\pi) } \Big\}\, .
\end{eqnarray}
where $k_1 \equiv p_1 - p_3$, $k_2 \equiv p_2 - p_4$, $l_1 \equiv p_1 - p_4$,
and $l_2 \equiv p_2 - p_3$, with $k_1 + k_2 = l_1 + l_2 = q$.
However, with $q^2 \approx m^2_h$, $k^2_1 \simeq - |\vk|^2$ and similarly for $k^2_2$,
$l^2_1$, and $l^2_2$, this contribution is subdominant.

The volume energy loss rate is
\begin{eqnarray}
\label{eq:Q_formula}
    Q_{N N \rightarrow N N \al \al} &=& \frac{\mathcal{S}}{2!} 
    \int \frac{d^3 \vec{q_1}}{2 \omega_1\, (2 \pi)^3}
    \frac{d^3 \vec{q_2}}{2 \omega_2\, (2 \pi)^3}\, \int \prod^{4}_{i=1}
    \frac{d^3 \vec{p_i}}{2 E_i\, (2 \pi)^3}\, f_1 f_2 (1-f_3) (1-f_4) \nonumber \\
    && \hspace{-2cm} \times\, \sum_{\rm spins} 
    |\mathcal{M}_{N N \rightarrow N N \al \al}|^2\, (2 \pi)^4 
    \delta^4 (p_1 + p_2 - p_3 - p_4 - q_1 - q_2)\, (\omega_1 + \omega_2)\, ,
\end{eqnarray}
where $\omega_1, \omega_2$ are the energy of the Goldstone bosons in the final state.
The symmetry factor ${\mathcal S}$ is $\frac{1}{4}$ for $n n $ and $p p$ 
interactions, whereas for $n p$ interactions it is $1$.
The nucleon occupation numbers are $f_i = 1 / (e^{(E_i - \mu_N) / T} + 1)$, 
where in the non-relativistic limit the nucleon energies are 
\begin{equation}
   E_i \simeq m_N + \frac{|\vec{p}_i|^2}{2 m_N} + U_N\, .
\end{equation}
Here $\mu_N$ is the chemical potential of the nucleon, and $U_N$ is the mean-field 
single-particle potential in which the nucleons move.
In Ref.~\cite{MartinezPinedo:2012rb} it is pointed out that due to the extreme 
neutron-rich conditions in the PNS core, the mean-field potentials for neutrons
and protons can differ significantly, with the difference directly related to
the nuclear symmetry energy (see e.g. Refs.~\cite{Baldo:2016jhp,Trautmann:2016ntm} 
for recent reviews).
Non-zero $U_n - U_p$ was found therein to have a strong impact on the spectra and
luminosities of the supernova emitted neutrinos.
In any case the nucleon occupation numbers are normalised to the nucleon
number density, 
\begin{equation}
\label{eq:occupation_number}
   n_N = X_N\, n_B = \int^\infty_0 \frac{2\, d^3 \vec{p}_i}{(2 \pi)^3}\, 
   f_i (\vec{p}_i)\, , 
\end{equation}
where $X_N$ with $N = n$, $p$, are the neutron and the proton fraction, respectively.
The relative abundances of the neutrons, protons, electrons, and the neutrinos
in the PNS core are determined by the conditions of kinetic and chemical equilibrium,
as well as charge neutrality. 
Therefore the neutron fraction $X_n$ parametrises the underlying nuclear equation of 
state and indicates the level of neutron degeneracy.

We perform the integral over the Goldstone boson momenta first
\begin{equation}
\label{eq:integral_GBmomenta}
   \hspace{-0.5cm} \int \frac{d^3 \vec{q_1}}{\omega_1}\, 
   \frac{d^3 \vec{q_2}}{\omega_2}\,
   \frac{(q_1 \cdot q_2)^2}{(q^2 - m^2_h)^2 + m^2_h \Gamma^2_h}\, 
   \frac{(2 q^2)^2}{(2 p \cdot q)^4}\, \omega 
   = \frac{2 (2 \pi)^2}{m^4_N} \int^\infty_0 d \omega\, 
   \omega^4\, I_1 (\omega, m_h, \vevr)\, ,         
\end{equation}
where $\omega = \omega_1 + \omega_2$.
The dimensionless integral is defined by
\begin{equation}
\label{eq:NNresonance}
   I_1 (\omega, m_h, \vevr) \equiv \int^1_0 d \tilde{\omega} 
   \int^{+1}_{-1} 
   \frac{d \cos \theta\, \tilom^5\, (1-\tilom)^5\, (1-\cos\theta)^4}{[2 \tilom\, 
   (1-\tilom)\, (1-\cos\theta) - \frac{m^2_h}{\omega^2}]^2 + \frac{m^2_h\, 
   \Gamma^2_h}{\omega^4}}\, ,
\end{equation}
with $\tilom \equiv \omega_1 / \omega$, and $\theta$ is the angle between the two 
emitted Goldstone bosons.

As the integral over the nucleon momenta in Eq.~(\ref{eq:Q_formula}) is not easy to evaluate, we follow the conventional approach of taking the non-degenerate and the degenerate limit in the following.
As we will show, energy loss rate due to Goldstone boson emission calculated in these
two limits have distinct dependences on the PNS core temperature $T$ and neutron
fraction $X_n$ therein.

\subsubsection{Non-degenerate limit}

The initial-state nucleon occupation numbers are given by
the non-relativistic Maxwell-Boltzmann distribution
$f_i (\vec{p}_i) = (n_N / 2) (2 \pi / m_N T)^{3/2} e^{- \vert \vec{p}_i 
\vert^2 / 2 m_N T}$.
The integration is simplified by introducing the center-of-mass momenta, so that
$\vec{p}_{1, 2} = \vec{P} \pm \vec{p}_i$ ,and 
$\vec{p}_{3, 4} = \vec{P} \pm \vec{p}_f$.
The $d^3 \vec{P}$ integral can be performed separately.
The energy loss rate in the non-degenerate limit is then
\begin{equation}
   Q^{\rm OPE\, (ND)}_{N N \rightarrow N N \al \al} =  
   \frac{\mathcal{S} \sqrt{\pi}}{(2 \pi)^6}\, (3- \frac{2 \beta}{3})\, 
   I_0\, n^2_N\, \left(\frac{f_N g\, m_N}{m^2_H}\right)^2\, 
   \left(\frac{2 m_N f_\pi}{m_\pi}\right)^4 \cdot \frac{T^{5.5}}{m^{4.5}_N}\, .
\end{equation}
Here we have defined the integral $I_0$ by
\begin{equation}
   I_0 (T, m_h, \vevr) \equiv \int du\, dv\, dx\, x^4\, I_1 (x, T, m_h, \vevr)\, 
   \sqrt{uv}\, e^{-u}\, \delta (u - v - x)\, ,   
\end{equation}
with $u \equiv |\vec{p}_i|^2 / m_N T$, $v \equiv |\vec{p}_f|^2 / m_N T$, and
$x \equiv \omega / T$.
The $\beta$ term is 
\begin{equation}
   \beta \equiv \frac{3}{I_0} \int du\, dv\, dx\, x^4\, I_1 (x, m_h, \vevr)\, 
   \sqrt{uv}\, e^{-u}\, \delta (u - v - x) \int^{+1}_{-1} \frac{d z}{2}
   \frac{\vert \vk \cdot \vl \vert^2}{\vert \vk \vert^2 \vert \vl \vert^2}\, ,
\end{equation}
where $z \equiv \left(\vec{p}_i \cdot \vec{p}_f \right) / |\vec{p}_i| |\vec{p}_f|$,
the angle between $\vec{p}_i$ and $\vec{p}_f$.

In the resonance region, one can make use of the limit of the Poisson kernel 
\begin{equation}
\label{eq:Poissonkernel}
   \lim_{\epsilon \rightarrow 0} \frac{1}{\pi} \frac{\epsilon}{a^2 + \epsilon^2} 
   = \delta (a)\, ,
\end{equation}
and obtain
\begin{equation}
\label{eq:PkI1}
   I^{\rm Pk}_1 (\omega, m_h, \vevr) \approx 
   \frac{\pi}{32} \frac{m^7_h}{\Gamma_h\, \omega^6}\, .
\end{equation}
Since this approximation is valid when $m^2_h / \omega^2 \approx 2 
\tilom \left(1 - \tilom \right)$, where the latter $\leq 1$, it is only applicable
for $\omega \geq m_h$ and $\Gamma_h \ll \omega$.
We have checked that, for $m_h = 500~\Mev$ and $\vevr = 10~\Gev$, this approximation 
still works well.

This is equivalent to considering the production of a real light Higgs boson $h$,
for which
\begin{eqnarray}
   Q^{\rm OPE\,(ND)}_{N N \rightarrow N N h} &=& \frac{\mathcal{S}\, 
   \sqrt{\pi}}{4\, \twopi^4} \left(3 - \frac{2 \beta}{3} \right)\, n^2_N 
   \left(\frac{f_N\, g\, \vevr\, m_N}{m^2_H} \right)^2
   \left(\frac{2 m_N f_\pi}{m_\pi} \right)^4\, 
   \frac{m^4_h}{m^{9/2}_N\, T^{1/2}} \nonumber \\
   && \times \int^\infty_{m_h / T} d x \frac{\sqrt{x^2 - \frac{m^2_h}{T^2}}}{x^3}
   \int^\infty_0 d u\, d v\, \sqrt{u v}\, e^{-u}\, \delta (u - v - x)\, . 
\end{eqnarray}
And indeed we find that for $m_h \lesssim 500~\Mev$,
\begin{equation}
   Q^{\rm P k} _{N N \rightarrow N N \al \al} \approx
   Q_{N N \rightarrow N N h}\, \times \mathcal{B} (h \rightarrow \al \al)\, ,
\end{equation}
with $\mathcal{B} (h \rightarrow \al \al) 
= \Gamma_{h \rightarrow \al \al} / \Gamma_h$ the branching ratio of the light Higgs 
boson $h$ decaying into a pair of Goldstone bosons.
Thus we find that in the parameter range we consider in this work, Goldstone boson production in the PNS core is dominated by the production of a real
light Higgs boson $h$ and its subsequent decay.
This is a very distinct feature from the nuclear bremsstrahlung emission of a 
massless scalar, e.g. the dilaton~\cite{Ishizuka:1989ts}, or a massive stable
scalar such as the saxion~\cite{Arndt:2002yg}.

\subsubsection{Degenerate limit}

We calculate the energy loss rate due to Goldstone boson emission, Eq.~(\ref{eq:Q_formula}), in the degenerate limit following Ref.~\cite{Friman:1978zq}.
The integral over the Goldstone boson momenta is done as in 
Eq.~(\ref{eq:integral_GBmomenta}) and Eq.~(\ref{eq:NNresonance}) first.
In the degenerate limit, the nucleon momenta integral is simplified by
$d^3 \vec{p}_i = |\vec{p}_j|^2 d |\vec{p}_i|\, d \Omega_i \approx
p_{\rm F} (n)\, m_N d E_j$.
The neutron Fermi momentum is $p_{\rm F} (n) = \left(3 \pi^2 n_n \right)^{1/3}$, 
with the neutron number density $n_n = X_n \rho / m_N$ given by 
Eq.~(\ref{eq:occupation_number}).
One then perform the integral
\begin{eqnarray}
\label{eq:FNN}
   \left< F_{N N} \right> &\equiv& \frac{(4 \pi)^2}{A} \int \prod^4_{i=1} 
   d \Omega_i\, \delta^3 (\vec{p}_1 + \vec{p}_2 - \vec{p}_3 - \vec{p}_4) 
   \times \\ \nonumber 
   && \Big\{\frac{\vert \vk \vert^4}{(\vert \vk \vert^2 + m^2_\pi)^2} + 
   \frac{\vert \vl \vert^4}{(\vert \vl \vert^2 + m^2_\pi)^2} 
   + \frac{\vert \vk \vert^2 \vert \vl \vert^2 - 2 |\vk \cdot \vl|^2}
    {(\vert \vk \vert^2 + m^2_\pi) (\vert \vl \vert^2 + m^2_\pi)} + 
    ... \Big\} \\ \nonumber
    &=& 3 - 5 x \tan^{-1} \left( \frac{1}{x} \right) +
   \frac{x^2}{1 + x^2} + \frac{x^2}{\sqrt{1 + 2 x^2}} \tan^{-1} 
   \left( \frac{\sqrt{1 + 2 x^2}}{x^2} \right)\, ,
\end{eqnarray}
with $A = (4 \pi)^5 / 2 p^3_{\rm F} (n)$, and $x \equiv m_\pi / 2 p_F (n)$.
The level of nucleon degeneracy is characterised by the $|\vk \cdot \vl|^2$ term.
In the case of strong degeneracy, $|\vk \cdot \vl|^2 = 0$.
Note also that in the degenerate limit, the pion mass terms $m^2_\pi$ in the braces
cannot be neglected.
Finally performing the integral over the nucleon energies yields
\begin{equation}
   \int \prod^4_{i=1} d E_i\, f_1\, f_2\, (1 - f_3)\, (1 - f_4)\, 
   \delta (E_1 + E_2 - E_3 - E_4 - \omega) = T^3\, J_{\al \al} (y)\, , 
\end{equation}
with $y \equiv \omega / T$, and
\begin{equation}
   J_{\al \al} (y) = - \frac{1}{6} \left(y^3 + 4 \pi^2 y \right)\, 
   \left(1 - e^y \right)^{-1}\, . 
\end{equation}

\begin{center}
\begin{figure}[t!]
\includegraphics[width=0.6\textwidth,angle=-90]{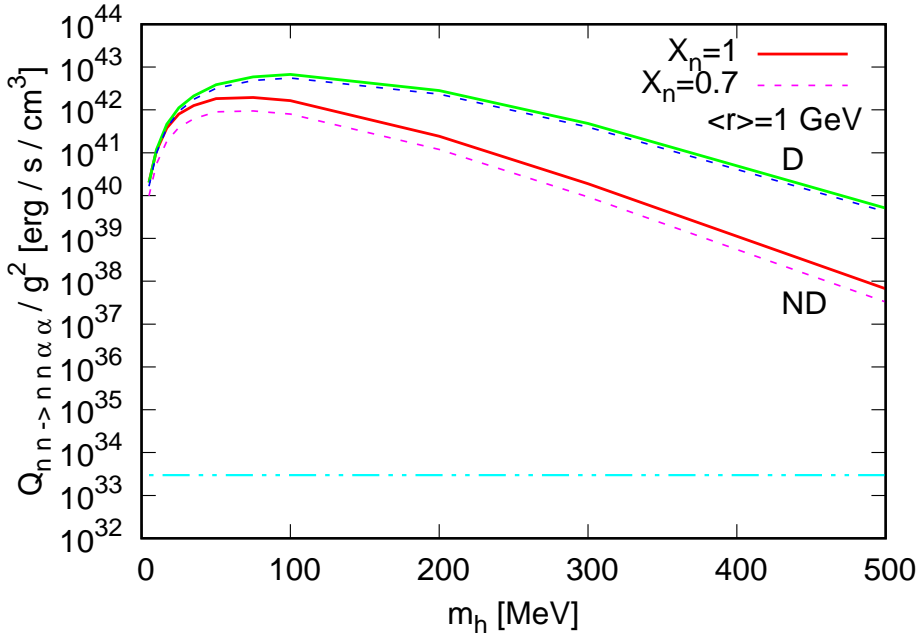}
\caption{Energy loss rate due to Goldstone boson emission from nuclear
bremsstrahlung processes $n n \rightarrow n n \al \al$ divided by the Higgs
portal coupling $g^2$, for various light Higgs boson mass $m_h$.
The rates are calculated in the non-degenerate (ND) and degenerate (D) limits, 
for proto-neutron star core temperature $T = 30~{\rm MeV}$, 
neutron fraction $X_n = 1$ (solid) and $0.7$ (dashed), respectively.
For all $m_h$ values we assume the radial field vacuum expectation value is
$\vevr = 1~\Gev$.
Also shown is Raffelt's analytical criterion on the energy loss rate per unit
volume $Q_X$ in Eq.~(\ref{eq:emissivity_bound}) (dash-double-dotted).}
\label{fig:GBemiss}
\end{figure}
\end{center}

\begin{center}
\begin{figure}[t!]
\includegraphics[width=0.6\textwidth,angle=-90]{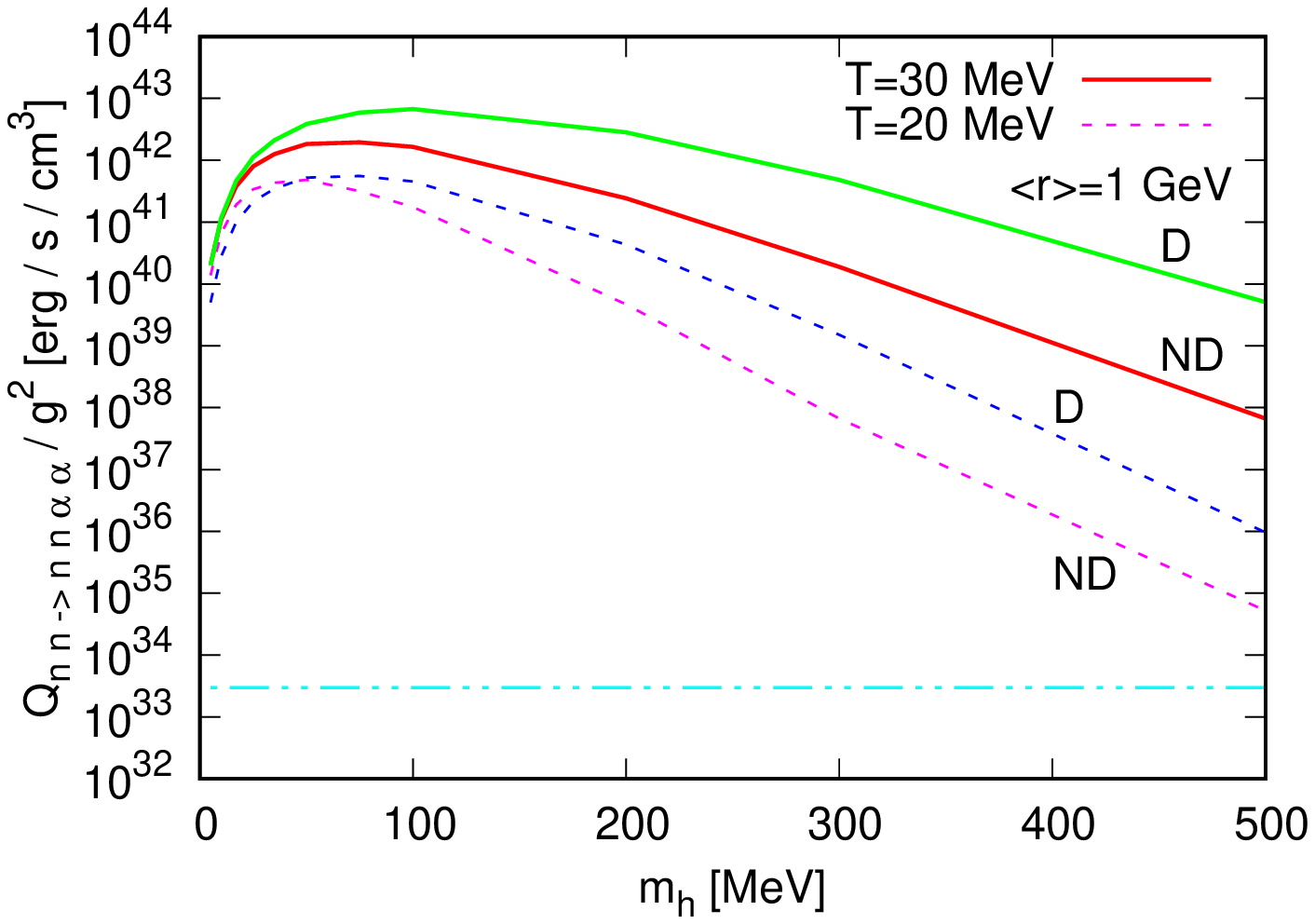}
\caption{Energy loss rate due to Goldstone boson emission from nuclear
bremsstrahlung processes $n n \rightarrow n n \al \al$ divided by the Higgs
portal coupling $g^2$, for various light Higgs boson mass $m_h$.
The rates are calculated in the non-degenerate (ND) and degenerate (D) limits, 
for proto-neutron star core temperature $T = 30~\Mev$ (solid) and $20~{\rm MeV}$ (dashed), and neutron fraction $X_n = 1$.
For all $m_h$ values we assume the radial field vacuum expectation value is
$\vevr = 1~\Gev$.
Also shown is Raffelt's analytical criterion on the energy loss rate per unit
volume $Q_X$ in Eq.~(\ref{eq:emissivity_bound}) (dash-double-dotted).}
\label{fig:GBemissT20}
\end{figure}
\end{center}

The energy loss rate in the degenerate limit is then
\begin{equation}
   Q^{\rm OPE\, (D)}_{N N \rightarrow N N \al \al} = 
   \frac{\mathcal{S}}{\left(2\pi \right)^9}\, 
   4 \left< F_{N N} \right>\, I_{\al \al}\, 
   \left(\frac{f_N\, g\, m_N}{m^2_H} \right)^2 
   \left(\frac{2 m_N f_\pi}{m_\pi} \right)^4\, p_F (n)\, \frac{T^8}{m^2_N}\, ,
\end{equation}
with the function given by
\begin{equation}
   I_{\al \al} (m_h, \vevr) \equiv \int^\infty_0 d y\, y^4\, I_1 (y, m_h, \vevr)\, 
   J_{\al \al} (y)\, . 
\end{equation}
We evaluate $I_{\al \al}$ numerically using the VEGAS subroutine both directly and 
using the limit of the Poisson kernel, Eq.~(\ref{eq:PkI1}).
Here we also checked that Goldstone boson production can be well described by the
production of a real light Higgs boson and its subsequent decay.
We compare the results in these two limits at the nuclear saturation density 
$\rho = 3 \cdot 10^{14}~{\rm g}/{\rm cm}^3$. 
In Fig.~\ref{fig:GBemiss} the comparison is made at the PNS core 
temperature $T = 30~\Mev$ and neutron fraction $X_n = 1$ and $0.7$.
Energy loss rate calculated in the two limits have different dependence on $X_n$:
$Q^{\rm (ND)}_{N N \rightarrow N N \al \al} \propto X^2_n$, and 
$Q^{\rm (D)}_{N N \rightarrow N N \al \al} \propto X^{1/3}_n$.
In Fig.~\ref{fig:GBemissT20} the comparison is made at two different PNS core
temperature $T = 30~\Mev$ and $20~\Mev$.

It was pointed out that in the case of a mixture of neutrons and protons, in the 
degenerate limit the energy loss rate for
$n p \rightarrow n p \al \al$ dominates that for $n n \rightarrow n n \al \al$
and $p p \rightarrow p p \al \al$, for all lepton fraction $Y_p$ values.
In Ref.~\cite{Brinkmann:1988vi} the axion emission rate was evaluated numerically
for arbitrary neutron degeneracies. 
It was found therein that the non-degenerate, analytical rate is a very good 
approximation. 
More recently, neutrino processes in post-collapse supernova core was studied in the partially-degenerate regime in Ref.~\cite{Bacca:2011qd}.
In this work we consider $n n$ interactions with $X_n = 1$ in the non-degenerate 
limit.

\subsection{Energy loss rate using phase shifts data}

One can also use the experimentally measured cross sections for $N N$ elastic 
scattering to obtain amplitude estimates for the nuclear bremsstrahlung 
processes.
Many independent observables are available from the nucleon-nucleon elastic 
scattering data collected by the EDDA Experiment at the Cooler Synchrotron (COSY) in 
J\"ulich~\cite{Albers:2004iw,Wilkin:2016qio}, experiments at the SATURNE II accelerator 
at Saclay, at the PSI, Ohio University, JINR, TSL in Uppsala, TUNL etc. 
(see e.g. Ref.~\cite{Arndt:2000xc,Arndt:2007qn}.)
In $NN$ interactions, the values of the total spin $\vec{S}$ and total angular 
momentum $\vec{J} = \vec{L} + \vec{S}$ are conserved, but that of the 
orbital angular momentum $\vec{L}$ may change because of the tensor force.
Therefore for $S = 1$, partial wave states $\ell_< = |J - 1|$ and 
$\ell_> = J + 1$ can couple to each other. 
In this case the scattering S-matrix has a $2 \times 2$ matrix structure, 
parametrised by the mixing angle $\epsilon_J$.
The diagonal elements are given by $e^{2 i\, \delta_{\ell_<}}\, \cos 2 \epsilon_J$ and
$e^{2 i\, \delta_{\ell_>}}\, \cos 2 \epsilon_J$, respectively, and
the off-diagonal elements are both 
$i\, e^{i\, \left(\delta_{\ell_>} + \delta_{\ell_>} \right)}\, \sin 2 \epsilon_J$.
Phase shifts $\delta_{\ell S J}$ and mixing angles 
$\epsilon_J$ for a wide range of laboratory kinetic energies $T_{\rm lab}$ are 
available at the Nijmegen NN-OnLine website~\cite{Nijmegen}.
Full data and a number of fits to data are available on the 
SAID database~\cite{SAID}. 
In the energy range below $25~{\rm MeV}$, there are numerous measurements on the
total $np$ cross section, but not on $pp$ due to the large Coulomb effects. 
Therefore the uncertainties in the latter are larger.

A nice summary of the general formalism for two-body scattering of spin-$1/2$ particles can be found in Ref.~\cite{Kang:2014ioa}.
The total cross section for $pp$ elastic 
scattering is simply
\begin{equation}
\label{eq:sigma_pwa}
  \sigma_{N N} = 2 \pi\, \sum_{J} (2 J + 1)\, |f_J (\vec{k}_{\rm cm} )|^2
  = \frac{2 \pi}{|\vec{k}_{\rm cm}|^2}\, \sum_{J} (2 J + 1)\, 
  \sin^2 \delta_{\ell S J}\, (\vec{k}_{\rm cm} )\, ,
\end{equation}
where $\vec{k}_{\rm cm}$ is the momentum in the centre-of-mass system, 
related to the laboratory kinetic energy as $|\vec{k}_{\rm cm}|^2 = \frac{1}{2} m_p\, T_{\rm lab}$, with $m_p$ the proton mass.

\subsubsection{Global fits of total elastic cross sections}

\begin{center}
\begin{figure}[t!]
\includegraphics[width=0.6\textwidth,angle=-90]{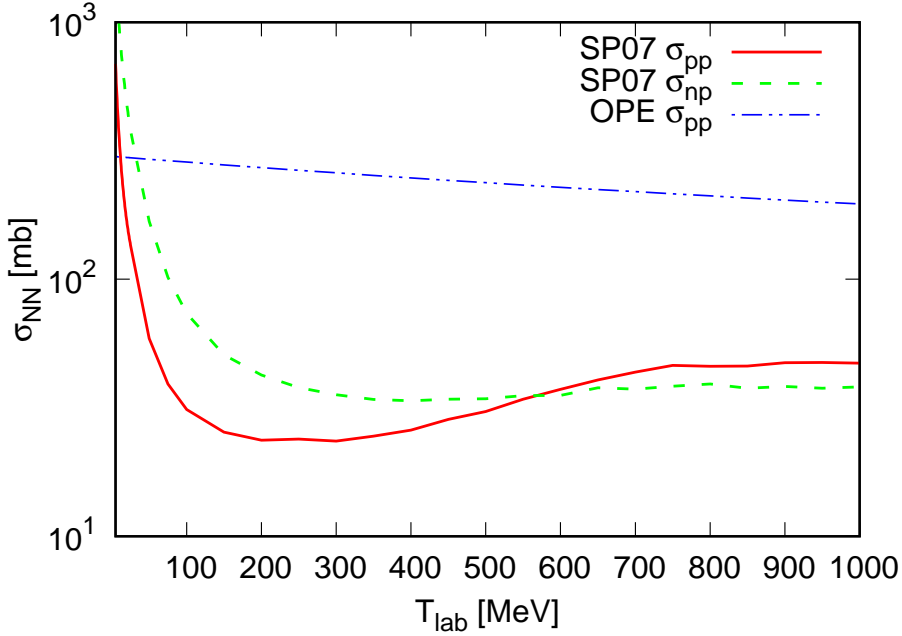}
\caption{The SP07 global fits for the total $pp$ (solid) and $np$ (dashed)
elastic scattering cross sections as a function of the laboratory kinetic energy 
$T_{\rm lab}$, reported in Ref.~\cite{Arndt:2008uc}.
Also plotted is the total $pp$ elastic cross section obtained using the one-pion
exchange (OPE) approach (dash-double-dotted), with the pion mass $m_\pi$ in the 
braces neglected.}
\label{fig:sigma_SP07LE08}
\end{figure}
\end{center}

In this work we use the SP07 and LE08 global fits for the total proton-proton and
neutron-proton elastic scattering cross sections $\sigma_{pp}$ and $\sigma_{np}$~\cite{Arndt:2007qn,Arndt:2008uc}, respectively, as shown in 
Fig.~\ref{fig:sigma_SP07LE08}.
The errors quoted therein are quite small, ranging from $0.01~{\rm mb}$ for low 
incident energies to $0.8~{\rm mb}$ at most for high incident energies.
The huge cross section at zero-energy indicates that there is a two-body bound
state, or quasi-bound state, as manifested in the negative scattering lengths
$a_{pp} \approx -17.1~{\rm fm}$ and $a_{n p} \approx -23.74~{\rm fm}$
(see e.g. Ref.~\cite{Entem:2017gor}.)
We also plot the $NN$ elastic scattering cross section calculated using the OPE
approximation, where for simplicity we neglect the pion mass $m_\pi$ in the 
the braces in the amplitude expression
\begin{eqnarray} 
\label{eq:MNN_OPE}
    \sum_{\rm spins} |\mathcal{M}^{\rm OPE}_{N N \rightarrow N N}|^2 &=&    
    4\, \left(\frac{2 m_N f_\pi}{m_\pi} \right)^4         
    \Big\{\frac{|\vk|^4}{(|\vk|^2 + m^2_\pi)^2} + 
    \frac{|\vl|^4}{(|\vl|^2 + m^2_\pi)^2} \nonumber \\ 
    && + \frac{|\vk|^2\, |\vl|^2 + 
    2 (\vk \cdot \vl)^2 - 2 (|\vk|^2 + |\vl|^2) (\vk \cdot \vl)}
    {(|\vk|^2 + m^2_\pi) (|\vl|^2 + m^2_\pi)} \Big\}\, .
\end{eqnarray}
As expected, the OPE approximation is good only for 
$T_{\rm lab} \simeq 10$--$20~\Mev$.
For larger laboratory kinetic energies, it overetimates by a factor of $10$ 
(for $T_{\rm lab} \simeq 100$--$400~\Mev$) to 
$4$ (for $T_{\rm lab} \simeq 800$--$1000~\Mev$).

Results in Ref.~\cite{Albers:2004iw} show that for low energy scattering, 
$d \sigma_{N N} / d \Omega$ has no strong angular dependence. 
Therefore we simply use
$\sum_{\rm spins} |\mathM_{N N}|^2 \approx 64\, |\mathA_{N N}|^2 m^4_N$ to infer
$|\mathA_{N N}|^2$ as a function of the center-of-mass energy
$E^2_{\rm cm} \approx 4 m^2_N + 2 m_N T_{\rm lab}$.

\begin{center}
\begin{figure}[t!]
\includegraphics[width=0.6\textwidth,angle=-90]{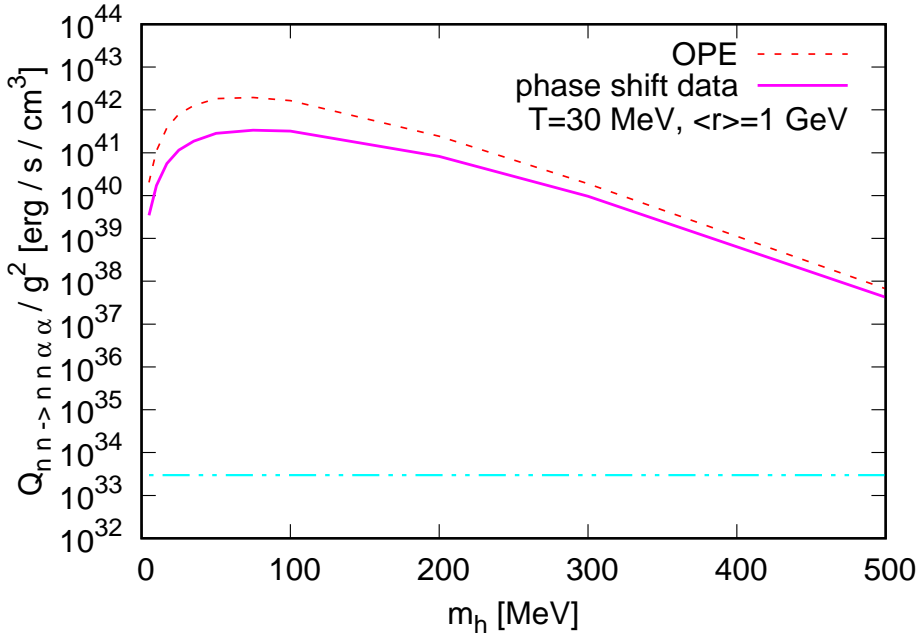}
\caption{Energy loss rate due to Goldstone boson emission from nuclear
bremsstrahlung processes $n n \rightarrow n n \al \al$ divided by the Higgs
portal coupling $g^2$, for various light Higgs boson mass $m_h$. 
The rates are calculated using the one-pion exchange (OPE) approximation (dashed) 
and the SP07 global fits for the total $pp$ elastic cross section (solid), and 
assume $\sigma_{nn} = \sigma_{pp}$.
Here we take the non-degenerate (ND) limit only, and set
proto-neutron star core temperature $T = 30~{\rm MeV}$, and neutron fraction 
$X_n = 1$.
For all $m_h$ values we assume the radial field vacuum expectation value is
$\vevr = 1~\Gev$.
Also shown is Raffelt's analytical criterion on the energy loss rate per unit
volume $Q_X$ in Eq.~(\ref{eq:emissivity_bound}) (dash-double-dotted).}
\label{fig:GBemissexp}
\end{figure}
\end{center}

With this information, we estimate the amplitude squared for the 
nuclear bremsstrahlung processes $N N \rightarrow N N \al \al$ 
\begin{equation}
   \sum_{\rm spins} |\mathcal{M}^{\rm exp}_{N N \rightarrow N N \al \al}|^2 \approx
   1024\, |\mathcal{A}_{N N}|^2\, \left(\frac{f_N\, g m_N}{m^2_H} \right)^2
   \frac{\left(q_1 \cdot q_2 \right)^2}{\left(q^2 - m^2_h \right)^2 + m^2_h 
   \Gamma^2_h} \frac{\left(-2 q^2 \right)^2}{\left(2 p \cdot q \right)^4}\,
   m^6_N\, ,
\end{equation}
after summing over 64 direct and exchange diagrams.
To evaluate the phase space integral in the energy loss rate calculation, we take the non-degenerate limit, and proceed as in the OPE case.
The energy loss rate is then
\begin{equation}
   Q^{\rm exp\, (ND)}_{N N \rightarrow N N \al \al} = 
   \frac{32\, \mathcal{S}}    {\twopi^6}\, 
   I^{exp}_0\, n^2_B \left(\frac{f_N\, g m_N }{m^2_H} \right)^2\, 
   \frac{T^{11/2}}{m^{1/2}_N}\, .
\end{equation}
Here we define the integral
\begin{eqnarray}
   I^{exp}_0 (T, m_h, \vevr) &\equiv& \int du\, dv\, dx\, dy\, x^4 
   I_1 (x, T, m_h, \vevr)\,
   \sqrt{y}\, e^{-y}\, \sqrt{u v}\, e^{-u}\, 
   \delta \left(u - v - x \right) \nonumber \\
   && \cdot\, |\mathA_{NN}|^2 \left (u, y \right)\, ,
\end{eqnarray}
with $y \equiv |\vec{P}|^2 / m_N T$.
The result obtained by using the SP07 global fit to the $\sigma_{pp}$ data
is plotted in Fig.~\ref{fig:GBemissexp} and compared to the OPE result. 
The overestimation by OPE happens to be milder for $N N \rightarrow N N \al \al$
than in $N N \rightarrow N N$, because of the different kinematics of the
exchanged pion in the nuclear bremsstrahlung processes from that in the elastic
scattering.

For neutrino emission from the $n n \rightarrow n n \nu \bar{\nu}$ processes,
Ref.~~\cite{Hanhart:2000ae} used on-shell $N N$ amplitudes measured by experiments
and found that the OPE approximation overestimated the energy loss rate by about 
a factor of four.
Ref.~\cite{Bacca:2015tva} found that the next-to-next-to-next-to-leading order 
(${\rm N^{3}LO}$) chiral effective field theory calculations differ by about a factor 2--3 from leading order (LO) results, and the result obtained by using the experimental phase shifts data is very similar to the ${\rm N^{3}LO}$ ones. 
Since the central contact terms in the chiral effective field theory do not contribute
in the nuclear bremsstrahlung processes, the leading-order term is solely the
one-pion exchange potential.
For axions, the OPE approximation is also found to oversimplify the nuclear dynamics 
and overestimate the emission rate by a factor of four~\cite{Hanhart:2000ae}.

\subsubsection{Chiral effective field theory predictions}

\begin{center}
\begin{figure}[t!]
\includegraphics[width=0.6\textwidth,angle=-90]{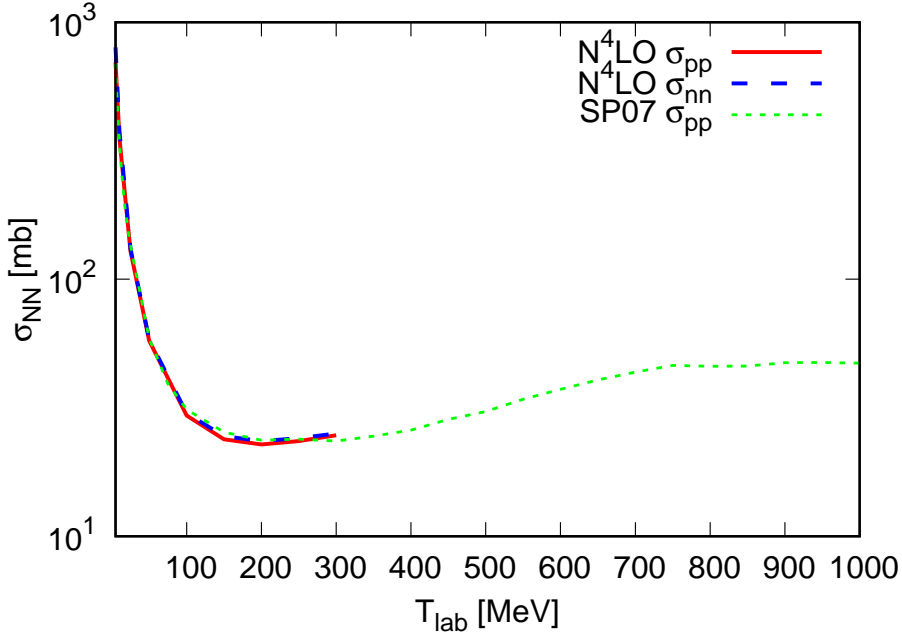}
\caption{Total $pp$ (solid) and $nn$ (dashed) elastic scattering cross sections as a function of the lab kinetic energy $T_{\rm lab}$, from the ${\rm N^4 LO}$ chiral effective field theory results for the phase shifts presented in Ref.~\cite{Entem:2017gor}.
Also plotted is the SP07 global fits for the total $pp$ elastic cross section (dotted)
reported in Ref.~\cite{Arndt:2008uc}.}
\label{fig:sigma_chiralEFT}
\end{figure}
\end{center}

Charge independence breaking (CIB) of the strong $NN$ interactions refers to the 
difference between the isospin $I=1$ states: the proton-proton ($I_z = +1$),
the neutron-proton ($I_z = 0$), and the neutron-neutron ($I_z = -1$) interactions,
after electromagnetic effects are removed.
Charge symmetry breaking (CSB) concerns the difference between the $pp$ and $nn$
interactions only.
CIB is clearly seen in Fig.~\ref{fig:sigma_SP07LE08}, while
a small amount of CSB is observed in the measured scattering lengths 
$a_{nn}$ and $a_{pp}$, as well as the effective range $r_{nn}$ and $r_{p p}$.
A detailed discussion on charge-dependence of nuclear interactions can be found in
Ref.~\cite{Machleidt:2011zz} (see also Ref.~\cite{Konobeevski:2017mpw}.)
Very recently, Ref.~\cite{Entem:2017gor} provides $pp$, $nn$ and $np$
phase shifts predicted by the chiral effective field theory to the ${\rm N^{4} LO}$.
In all partial waves, the predicted $n p$ phase shifts and mixing angles at this order 
are shown to agree excellently with the Nijmegen multi-energy~\cite{Stoks:1993tb} 
and the SP07 single-energy analysis~\cite{Arndt:2007qn}. 
Charge-dependence due to pion-mass splitting is taken into account in the one-pion
exchange terms only, while nucleon-mass splitting is always included. 
Fig.~\ref{fig:sigma_chiralEFT} shows total $pp$ and $nn$ elastic cross 
sections calculated with Eq.~(\ref{eq:sigma_pwa}) using
the ${\rm N^{4} LO}$ chiral effective field theory phase shifts from 
Ref.~\cite{Entem:2017gor}.
The $pp$ results agree very well with the SP07 global fit results.
For $T_{\rm lab} \lesssim 10~{\rm MeV}$, Coulomb force in $pp$ collisions is 
significant.
At larger laboratory kinetic energies, chiral effective field theory calculations 
predict that the effects of charge symmetry breaking is $\lesssim 3\%$ only.
In this work we therefore use the experimental data and set 
$\sigma_{nn} = \sigma_{pp}$.

Low-energy theorems~\cite{Low:1958sn,Adler:1966gc,Heller:1969ur} state that the 
first two terms in the series expansion of the bremsstrahlung amplitude
in powers of the energy loss may be exactly calculated by using the 
corresponding elastic, i.e. non-radiative, amplitude. 
In Ref.~\cite{Arndt:2002yg} it was argued that the model-independent approach 
of relating the nuclear bremsstrahlung amplitudes to the on-shell $N N$ 
scattering amplitudes measured by experiments is not applicable to scalar particles
such as the saxion.
The reason is that the contributions to the leading order terms 
($\propto \omega^{-1}$) from the emission 
of a scalar particle from external nucleon legs cancel each other, which does not
happen for axion and neutrino pairs~\cite{Hanhart:2000ae}, or 
KK-gravitons~\cite{Hanhart:2000er}.
The next-to-leading order term ($\propto \omega^0$) includes the 
emission diagrams of the scalar particle from external legs as well as from 
internal lines, where the latter is not calculable due to the unknown interaction vertices, and may be dominant. 

In Weinberg's Higgs portal model,
we also found the cancellation of the leading order terms between the 
diagrams for the Goldstone boson pairs being emitted from the external nucleon legs.
The effective Higgs-pion coupling is $\propto (q^2 + \frac{11}{2}\, m^2_\pi) / \vevh$, 
so the emission from internal lines is of order $\mathcal{O} (\omega^0)$ 
as well in the low-energy limit (cf. Eq.~(\ref{eq:MOPE_pion})).
However, in Weinberg's Higgs portal model Goldstone boson production in the 
PNS core is dominated by the emission of a real light Higgs boson in nuclear
bremsstrahlung processes and its subsequent decay.
Therefore for small light Higgs boson mass $m_h$ the low-energy theorems should still
be applicable.
This remains to be verified by using the chiral effective field theory to calculate
the emission of the light Higgs boson $h$ from the external nucleon legs as well
as from the internal lines.

\section{Goldstone boson propagation in proto-neutron star core}
\label{sec:GBmfp}

In the weakly-interacting regime, the Goldstone boson mean free path is set by the 
elastic scattering rate $R_{\al N \rightarrow \al N}$.
In the strongly-interacting regime, the absorption rate 
$R_{N N \al \al \rightarrow N N}$ may be comparable. 
The mean free path in the former case is 
$l_{\rm mfp} = (n_B \sigma_{\al N \rightarrow \al N})^{-1}$, 
while in the latter case, the mean free path against absorption is 
$l^{\rm absorb.}_{\rm mfp} = (n^2_B \sigma_{\al \al N N \rightarrow N N})^{-1}$.
For axions, Ref.~\cite{Turner:1987by} has considered the free-streaming regime, 
while Ref.~\cite{Burrows:1990pk} the trapping regime.

\begin{center}
\begin{figure}[t!]
\includegraphics[width=0.6\textwidth,angle=-90]{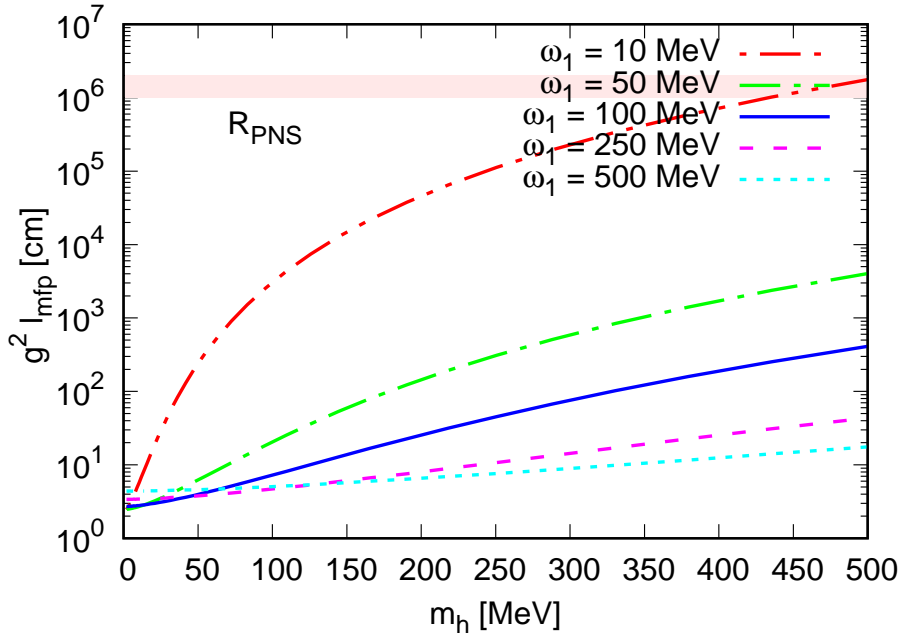}
\caption{Goldstone boson mean free path $l_{\rm mfp}$ times the Higgs portal
coupling $g^2$ in the proto-neutron star core versus the light Higgs boson mass $m_h$. 
Here we show the dependence on the incident Goldstone boson energy
for the values $\omega_1 = 10$ (dash double-dotted), 
$50$ (dash-dotted), $100$ (solid), $250$ (dashed), and $500~\Mev$ (dotted), 
respectively. 
Also shown is the proto-neutron star radius 
$R_{\rm PNS} \approx 10$--$20~{\rm km}$ (shaded region).}
\label{fig:GBmfp_omega1}
\end{figure}
\end{center}

The amplitude squared for the elastic process 
$\al (q_1) N (p_1) \rightarrow \al (q_2) N (p_1)$ is
\begin{equation}
   \sigma_{\al N \rightarrow \al N} = \frac{4 f^2_N g^2\, m^2_N}{m^4_\varphi}
   \frac{(q_1 \cdot q_2)^2\, \left[\left(p_1 \cdot p_2 \right) + m^2_N \right]}
   {(t - m^2_r)^2}\, .
\end{equation}
We follow Ref.~\cite{Tubbs:1975jx} to calculate the reaction rate 
\begin{eqnarray}
   R_{\al N \rightarrow \al N} &=& n_B\, \sigma_{\al N \rightarrow \al N}\, \vM 
   = \int \frac{2 d^3\vec{p}_1}{(2 \pi)^3} f (\vec{p}_1)\, 
   \frac{1}{2 \omega_1\, 2 E_1}\, 
   \int \frac{d^3\vec{q}_2}{(2 \pi)^3\, 2 \omega_2} \times \nonumber \\
   && \hspace{-2.5cm} \int \frac{d^3\vec{p}_2}{(2 \pi)^3\, 2 E_2}\,
   \left[1 - f (\vec{p}_2) \right]\,    
   \frac{1}{2} \sum_{\rm spins} |\mathcal{M}_{\al N \rightarrow \al N}|^2\, 
   (2 \pi)^4 \delta^4 (p_1 + q_1 - p_2 - q_2)\, .
\end{eqnarray}
Using the polar angle $\cos \theta \equiv \vec{p}_1 \cdot \vec{q}_1 / |\vec{p}_1| 
|\vec{q}_1|$ and the azimuthal angel $\phi^\prime$ which is measured from the 
$(\vec{p}_1, \vec{q}_1)$-plane, the 9-dimensional integral can be simplified to
\begin{eqnarray}   
\label{eq:RalNalN}
   R_{\al N \rightarrow \al N} &=& \frac{1}{(2 \pi)^3}\, \frac{m^4_N}{4 \omega_1} 
   \frac{f^2_N g^2 m^2_N}{m^4_\varphi} 
   \int^\infty_1 d \epsilon_1 f (\epsilon_1) \sqrt{\epsilon^2_1 - 1} 
   \int^{+1}_{-1} \frac{d \cos \theta}{\lambda (\epsilon_1, u_1, \cos \theta)} 
   \nonumber \\
   && \times \int^{\epsilon^{\rm max}_2}_{\epsilon^{\rm min}_2}  d \epsilon_2
   \left[1 - f (\epsilon_2) \right] \int^{2 \pi}_0 \frac{d \phi^\prime}{2 \pi}
   F_3\, ,
\end{eqnarray}
with the dimensionless variables 
$\epsilon_1 \equiv E_1 / m_N$, $\epsilon_2 \equiv E_2 / m_N$, 
and $u_1 \equiv \omega_1 / m_N$. 
The functions in the above equation are defined as
\begin{equation}
   \lambda (\epsilon_1, u_1, \cos \theta) \equiv \frac{|\vec{p}_1 + \vec{q}_1|}{m_N} 
   = \sqrt{\epsilon^2_1 - 1 + u^2_1 + 2 u_1 \left(\epsilon^2_1 - 1 \right)^{1/2} 
   \cos \theta}\, ,
\end{equation}
and
\begin{equation}
   F_3 \equiv \frac{\left[q_1 \cdot \left(p_1 + q_1 - p_2 \right) \right]^3
   + 2 m^2_N \left[q_1 \cdot \left(p_1 + q_1 - p_2 \right) \right]^2}
   {\left[ 2 q_1 \cdot \left(p_1 + q_1 - p_2 \right) + m^2_r \right]^2\, m^2_N}\, ,
\end{equation}
respectively, and the limits for the $d \epsilon_2$ integration are determined to be
\begin{equation}
   \epsilon^{\rm max , \, \min}_2 = \frac{1}{2}\, \left[\epsilon_1 + u_1 \pm
   \lambda (\epsilon_1, u_1, \cos \theta) + \frac{1}
   {\epsilon_1 + u_1 \pm \lambda (\epsilon_1, u_1, \cos \theta)} \right]\, .
\end{equation}
To evaluate $q_1 \cdot p_2$, we need to know the angle
\begin{equation}
   \cos \theta_{q_1 p_2} \equiv \cos \theta^\prime \cos \Delta_2 - 
   \sin \theta^\prime \sin \Delta_2 \cos \phi^\prime\, ,         
\end{equation}  
where
\begin{equation}
   \cos \Delta_1 = \frac{\sqrt{\epsilon^2_1 - 1} + u_1 \cos \theta}{\lambda}\, ,
   \hspace{0.4cm}
   \cos \Delta_2 = \frac{u_1 + \sqrt{\epsilon^2_1 - 1} \cos \theta}{\lambda}\, ,
\end{equation}
with $\Delta_1 + \Delta_2 = \theta$, and 
\begin{equation}
   \cos \theta^\prime = \frac{E_2 \left(E_1 + \omega_1 \right) - p_1 \cdot q_1
   - m^2_N}{|\vec{p_2}| |\vec{p_1} + \vec{q_1}|}\, .
\end{equation}
We evaluate Eq.~(\ref{eq:RalNalN}) numerically using the VEGAS subroutine.
For low incident Goldstone boson energies $\omega_1 \ll m_N$, the nuclear recoil 
effects can be neglected, and so the interaction rate can also be easily estimated by
\begin{eqnarray}
   R_{\al N \rightarrow \al N} &=& n_B\, \sigma_{\al N \rightarrow \al N}\, \vM
   \nonumber \\
   &=& n_B \frac{\omega^4_1}{16 \pi} \frac{f^2_N g^2}{m^4_\varphi} \int^{+1}_{-1} 
   d \cos \theta \frac{\omega^2_1 \left(1 - \cos \theta \right)^3 + 2 m^2_N
   \left(1 - \cos \theta \right)^2}{\left[2 \omega^2_1 \left(1 - \cos \theta \right)
   + m^2_r \right]^2}\, .
\end{eqnarray}
We found that the results from this method agree with those from the full calculation 
within $20\%$ for $\omega_1 \lesssim ~{\rm 100 MeV}$.
In Fig.~\ref{fig:GBmfp_omega1} we plot the Goldstone boson mean free path 
$l_{\rm map}$ times the Higgs portal coupling $g^2$ versus the
light Higgs boson mass $m_h$, for various incident Goldstone boson energies $\omega_1$.

\begin{center}
\begin{figure}[t!]
\includegraphics[width=0.6\textwidth,angle=-90]{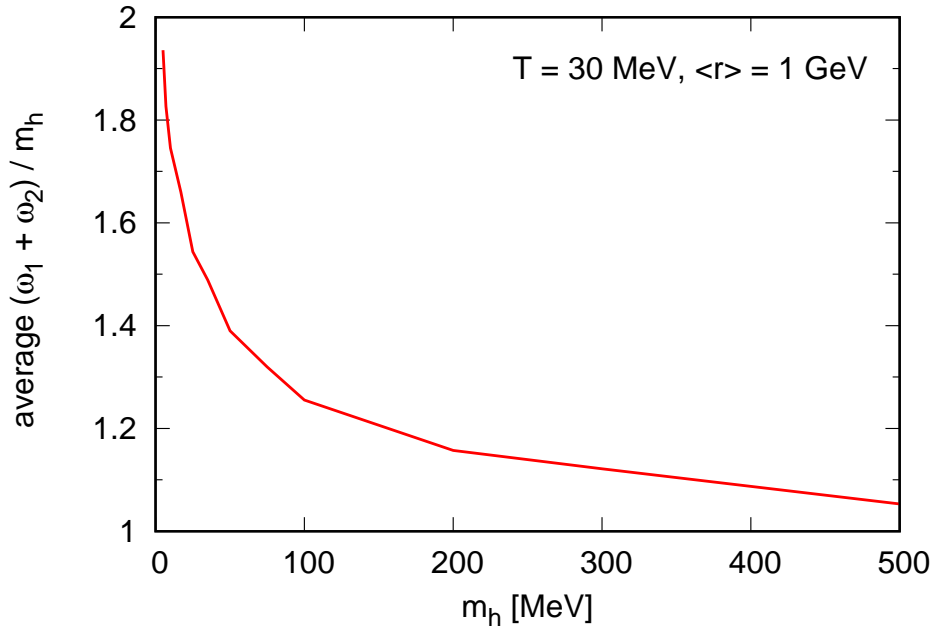}
\caption{Goldstone boson average emission energy 
$\overline{\omega} =\overline{\omega_1 + \omega_2}$ in dependence of the light Higgs 
boson mass $m_h$. 
Both the energy loss rate $Q_{n n \rightarrow n n \al \al}$ and the thermally 
averaged cross section 
$\left<\sigma_{n n \rightarrow n n \al \al} \right>\, v_{\rm M}$ are 
calculated in the non-degenerate (ND) limit, for proto-neutron star core 
temperature $T = 30~{\rm MeV}$, neutron fraction $X_n  =1$, 
and the radial field vacuum expectation value $\vevr = 1~{\rm GeV}$.}
\label{fig:GBavenergy}
\end{figure}
\end{center}

Goldstone boson pairs are emitted with an average energy of 
\begin{equation}
   \frac{\bar{\omega}}{T} = \frac{1}{T} 
   \frac{Q_{N N \rightarrow N N \al \al}}{n^2_N 
   \left< \sigma_{N N \rightarrow N N \al \al} v_{\rm M}\right>}\, ,
\end{equation}
where $v_{\rm M}$ is the M$\o$ller velocity.
In Fig.~\ref{fig:GBavenergy} we choose to plot the ratio of the Goldstone boson 
average emission energy to the light Higgs boson mass $m_h$. 
The curve indicates again that for $m_h \lesssim 500~\Mev$ Goldstone boson emission 
is still dominated by the production of a real light Higgs boson $h$.

\begin{center}
\begin{figure}[t!]
\includegraphics[width=0.6\textwidth,angle=-90]{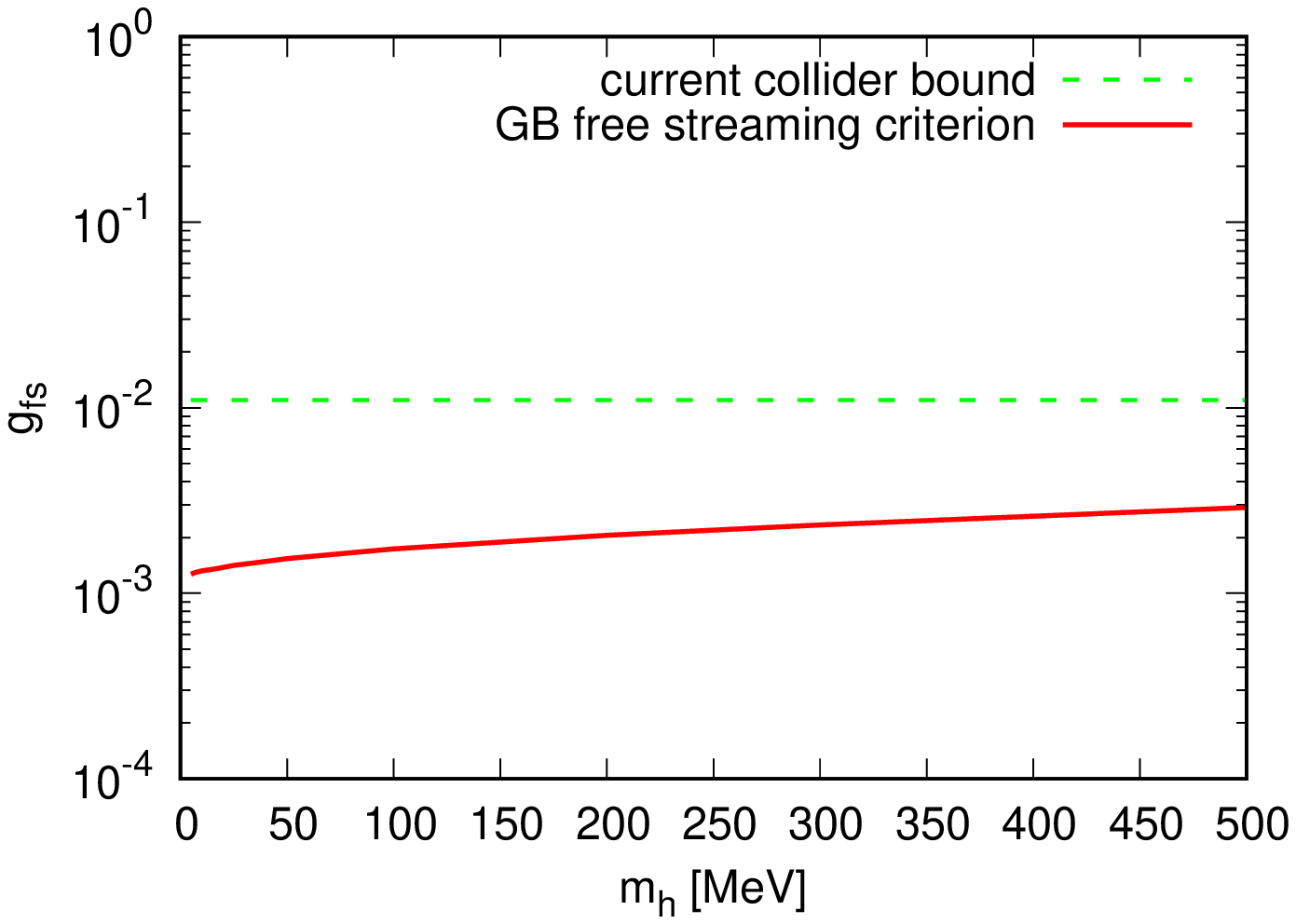}
\caption{Upper limits on the Higgs portal coupling $g$ for Goldstone boson 
free-streaming out of the proto-neutron star core, Eq.~(\ref{eq:freestreaming}),
for various light Higgs boson mass $m_h$ (solid).
Also shown is the current collider bound, Eq.~(\ref{eq:colliderbound}) (dashed).}
\label{fig:GBfreestreaming}
\end{figure}
\end{center}

We divide the free-streaming and the trapping regime by 
$l_{\rm mfp} \gg R_{\rm PNS}$ and $l_{\rm mfp} \ll R_{\rm PNS}$, respectively.
The neutron star radius is about $10~{\rm km}$~\cite{Guillot:2013wu,Raithel:2016vtt}, 
depending on the equation of state 
(see Refs.~\cite{Lattimer:2015nhk,Miller:2016pom} for recent reviews.)
But the proto-neutron star radius is about $10$--$20~{\rm km}$ at post-bounce times
$\lesssim 3~{\rm s}$, slightly larger than that of neutron stars, 
as shown in the simulations of e.g. Ref.~\cite{Fischer:2009af}.
Therefore, if the Higgs portal coupling saturates the collider bound $g \leq 0.011$, 
the Goldstone bosons would be trapped in the PNS core. 
In this case they still contribute to the cooling of the PNS core, and
one needs to estimate the opacity of the medium to the Goldstone bosons
as in Ref.~\cite{Burrows:1990pk} for axions. 
The amplitudes for the Goldstone boson pair absorption rate, 
$\sum_{\rm spins} |\mathcal{M}_{N N \al \al \rightarrow N N}|^2$, are the same as 
for the nuclear bremsstrahlung energy loss rate.
For simplicity, in this work we consider only the free-streaming regime by demanding  
\begin{equation}
\label{eq:freestreaming}
    g \lesssim \sqrt{\frac{g^2\, l_{\rm mfp} (\bar{\omega})}{R_{\rm PNS}}}
    \equiv g_{\rm fs}\, ,
\end{equation} 
for each light Higgs boson mass $m_h$.
We plot the Goldstone boson free-streaming criterion $g_{\rm fs}$ in 
Fig.~\ref{fig:GBfreestreaming}, assuming $R_{\rm PNS} = 20~{\rm km}$ for the 
proto-neutron star radius.
For $m_h \lesssim 50~\Mev$, it is beyond the projected sensitivity of future 
collider experiments for SM Higgs invisible decay (cf. Eq.~(\ref{eq:colliderbound})).

\section{SN 1987A constraints on Weinberg's Higgs portal model}
\label{sec:constraints}

\begin{center}
\begin{figure}[t!]
\includegraphics[width=0.6\textwidth,angle=-90]{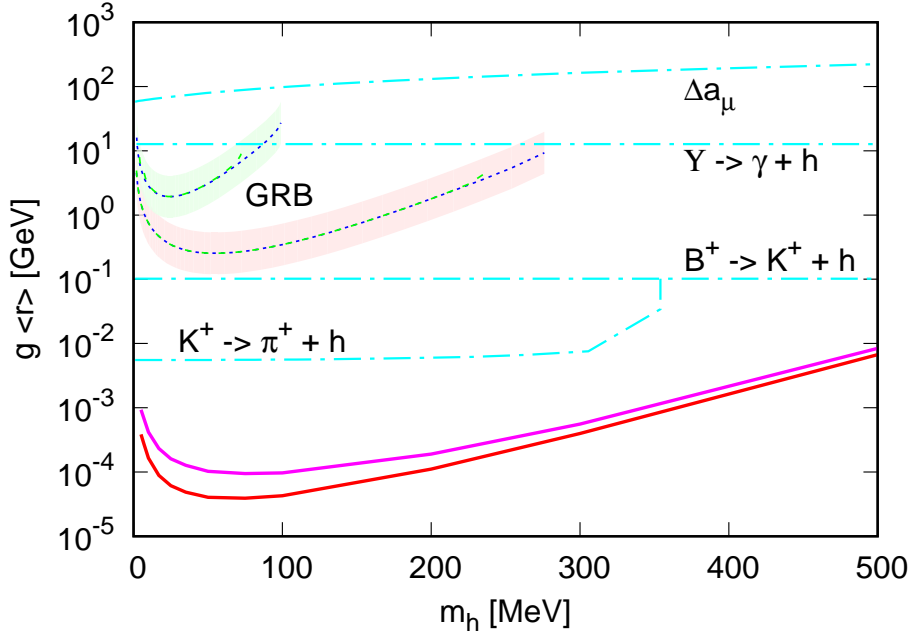}
\caption{SN 1987A upper limits on $g \vevr$, the product of the Higgs portal coupling
with the vacuum expectation value of the radial field $r$, for various light
Higgs boson mass $m_h$ (solid lines).
The upper solid curve is derived by using the SP07 global fits for the 
nucleon-nucleon elastic scattering cross section in the energy loss rate calculation, 
and the lower one by using the one-pion exchange (OPE) approximation.
Also shown are the upper limits set by laboratory experiments (dash-dotted lines, 
from top to bottom), such as the muon anomalous magnetic moment $\Delta a_\mu$, 
radiative Upsilon decays $\Upsilon (ns) \rightarrow \gamma + h$, $B^+$ meson
invisible decay $B^+ \rightarrow K^+ + h$, as well as $K^+$ meson invisible decay
$K^+ \rightarrow \pi^+ + h$.
The dotted and the dashed lines labelled "GRB" are the upper limits we derived in 
Ref.~\cite{Tu:2015lwv} by invoking the energy loss argument on the initial fireballs 
of gamma-ray bursts. 
Two GRB initial fireball temperatures values $T_0 = 18~\Mev$ (lower) and 
$8~\Mev$ (upper) are assumed, and the Higgs portal coupling $g$ is taken to 
saturate the current collider bound (dotted) and at future collider sensitivities
(dashed).
The uncertainties in these GRB upper limits resulting from the error in the GRB
energy loss argument, Eq.~(\ref{eq:Qcriterion}), are indicated by the filled 
regions.}
\label{fig:SNGRBmeson}
\end{figure}
\end{center}

Ideally one should do numerical simulations as in 
Refs.~\cite{Hanhart:2001fx,Keil:1996ju,Fischer:2016cyd} to study the effects of the additional cooling agent on the neutrino burst signal.
Here we simply invoke Raffelt's analytical 
criterion~\cite{Raffelt:1990yz,Raffelt:2006cw} on the energy loss rate per unit mass 
due to the emission of an exotic species $X$
\begin{equation}
\label{eq:emissivity_bound}
   \epsilon_X \equiv \frac{Q_X}{\rho} \lesssim 10^{19}\, {\rm erg} 
   \cdot {\rm g}^{-1} \cdot {\rm s}^{-1}\, ,
\end{equation}
as shown in Fig.~\ref{fig:GBemiss}, Fig.~\ref{fig:GBemissT20}, and 
Fig.~\ref{fig:GBemissexp}.
It is to be applied at typical PNS core conditions, i.e. at a temperature 
$T = 30~\Mev$, and baryon mass density 
$\rho = 3 \cdot 10^{14}~{\rm g} / {\rm cm}^3$.
The SN 1987A constraint on Weinberg's Higgs portal model is obtained by finding the 
model parameters $g$ and $\vevr$ for each light Higgs boson mass $m_h$ such that
the energy loss rate due to Goldstone boson emission 
$Q_{N N \rightarrow N N \al \al} < Q_X$. 
In the resonance region of producing a real light Higgs boson $h$, 
where the approximation with Poisson kernel limit is applicable, 
we have seen that $Q_{N N \rightarrow N N \al \al} \propto g \vevr$.
Therefore we scale the estimates for this quantity calculated using the one-pion 
exchange (OPE) approach and the SP07 global fits to the elastic $pp$ cross section, 
both in the non-degenerate (ND) limit, and assuming $\sigma_{nn} = \sigma_{pp}$ 
(cf. Fig.~\ref{fig:GBemissexp}) to be below $Q_X$.
Our main results are presented in Fig.~\ref{fig:SNGRBmeson}.
In these SN 1987A constraints, the collider bound and the free-streaming criterion on 
$g$ (Eq.~(\ref{eq:colliderbound}) and Eq.~(\ref{eq:freestreaming}), 
respectively), as well as the perturbativity condition on 
$\vevr$ (Eq.~(\ref{eq:perturbativity})) are all satisfied.  
We find that using OPE and the SP07 global fits results only in a factor 
of $2.6$ difference for $10~\Mev \lesssim m_h \lesssim 50~\Mev$, and a factor of 
$1.4$ for $m_h > 300~\Mev$.
Uncertainty from the effective Higgs-nucleon coupling $f_N$ is $\sim 10\%$.
We have not included Goldstone boson production from $np$ bremsstrahlung processes,
which would strengthen both bounds.
Quantifying and discussing many-body and medium effects, or the impact of nucleon
effective masses~\cite{Baldo:2014yja} in nuclear interactions are beyond the scope of this work.

Nevertheless, Fig.~\ref{fig:SNGRBmeson} makes clear that
with nuclear uncertainties taken into account, the SN 1987A constraints 
still surpass those set by laboratory experiments~\cite{Anchordoqui:2013bfa}, 
or by energy loss argument in other astrophysical objects~\cite{Tu:2015lwv}, 
which we briefly summarise below.
As first pointed out in Ref.~\cite{Bird:2004ts}, decays of $B$ mesons to $K$ mesons
plus missing energy can be an efficient probe of GeV or sub-GeV scalar dark matter.
In Refs.~\cite{Anchordoqui:2013bfa,Huang:2013oua} this consideration has been 
applied to Weinberg's Higgs portal model.
If the light Higgs boson is lighter than $354~\Mev$, the decay of $K$ meson to 
a pion plus missing energy is a more powerful probe.
We follow Ref.~\cite{Anchordoqui:2013bfa} and use the most stringent constraint
on the decay branching ratios,
\begin{equation}
   \mathcal{B} (B^+ \rightarrow K^+ + h) < 10^{-5}\, ,
\end{equation}
by the BaBar experiment~\cite{delAmoSanchez:2010bk}, and
\begin{equation}
   \mathcal{B} (K^+ \rightarrow \pi^+ + h) < 10^{-10}\, ,
\end{equation}
by the E787 and E949 experiments~\cite{Artamonov:2009sz} at the 
Brookhaven National Laboratory.
The former imposes a constraint on the $\varphi - r$ mixing angle 
(Eq.~(\ref{eq:mixingangle})) that 
$\theta < 0.0016$, for $m_h < m_B - m_K$, while the latter
$\theta < 8.7 \cdot 10^{-5}$, for $m_h < m_K - m_\pi = 354~{\rm MeV}$.
Recently, the LHCb Collaboration has published upper limits on the branching
fraction $\mathcal{B} (B^+ \rightarrow K^+ X) \times 
\mathcal{B} (X \rightarrow \mu^+ \mu^-)$, where $X$ is a hypothetical long-lived 
scalar particle~\cite{Aaij:2016qsm}. 
The limits at the $95\%$ confidence level vary between $2 \cdot 10^{-10}$ and 
$10^{-7}$, for the scalar particle mass in the range $250~\Mev < m (X) < 4700~\Mev$ 
and lifetime in the range $0.1~{\rm ps} < \tau (X) < 1000~{\rm ps}$.
However, since in Weinberg's Higgs portal model we find  
$\mathcal{B} (h \rightarrow \mu^+ \mu^-) \lesssim 10^{-12}$, the LHCb upper limits
are not applicable.
Also shown in Fig.~\ref{fig:SNGRBmeson} are exclusion curves derived using 
radiative Upsilon decays, 
$\mathcal{B} (\Upsilon (nS) \rightarrow \gamma + h) < 3 \cdot 10^{-6}$, as well
as muon anomalous magnetic moment, $\Delta a_\mu = 288 \cdot 10^{-11}$.
Neither of them is useful to constrain $g \vevr$.

In our previous work~\cite{Tu:2015lwv} we have derived constraints using 
gamma-ray bursts (GRB) observations.
Due to resonance effects, Goldstone boson pairs can be rapidly produced 
by electron-positron annihilation process in the initial fireballs of the GRBs.
On the other hand, the mean free path of the Goldstone bosons is larger than the 
size of the GRB initial fireballs, so they are not coupled to the GRB's relativistic
flow and can lead to significant energy loss.
Our GRB energy loss criterion is 
\begin{equation} 
\label{eq:Qcriterion}
   Q_{e^+ e^- \rightarrow \al \al}\, \Delta t^\prime 
   \approx Q_{e^+ e^- \rightarrow \al \al}\, \frac{1}{\Gamma_0} 
   \frac{\Delta R_0}{\beta_0} \gtrsim \frac{\mathcal{E}}{\Gamma_0 V_0}\, ,
\end{equation}
where we used generic values for the GRB initial fireballs, such as total energy 
$\mathcal{E} = 10^{52}~{\rm erg}$, temperature $T_0 = 18~{\rm MeV}$ as well as
$8~{\rm MeV}$, radius $R_0 = 10^{6.5}~{\rm cm}$, wind velocity
$\beta_0 = 1 /\sqrt{3}$, and the Lorentz factor is
$\Gamma_0 = 1 / \sqrt{1 - \beta^2_0}$.
In fact, the GRB bounds on $g \vevr$ have a slight dependence on the Higgs portal
coupling $g$, which becomes visible when the light Higgs boson decay braching ratio
to a pair of SM fermions, $\Gamma_{h \rightarrow f \bar{f}}$, is no longer 
negligible compared to that to a pair of Goldstone bosons, 
$\Gamma_{h \rightarrow \al \al }$.
We therefore considered $g = 0.011$ saturating the current collider bounds, as well
as $g = 0.0015$ which might be probed by future collider experiments.
The region bounded by the two GRB exclusion curves, including the filled regions 
around them, represents the parameter space in Weinberg's Higgs portal model that can
be probed by GRB physics.
The GRB bounds are subject to large uncertainties, and are much weaker than the 
SN 1987A constraints.
However, they are competitive to current laboratory constraints in the mass range of
$m_h / T_0 \lesssim 10$--$15$.
We conclude here that Weinberg's Higgs portal model is another example to elucidate
that high-energy astrophysical objects are excellent laboratory for particle
physics.

\begin{center}
\begin{figure}[t!]
\includegraphics[width=0.6\textwidth,angle=-90]{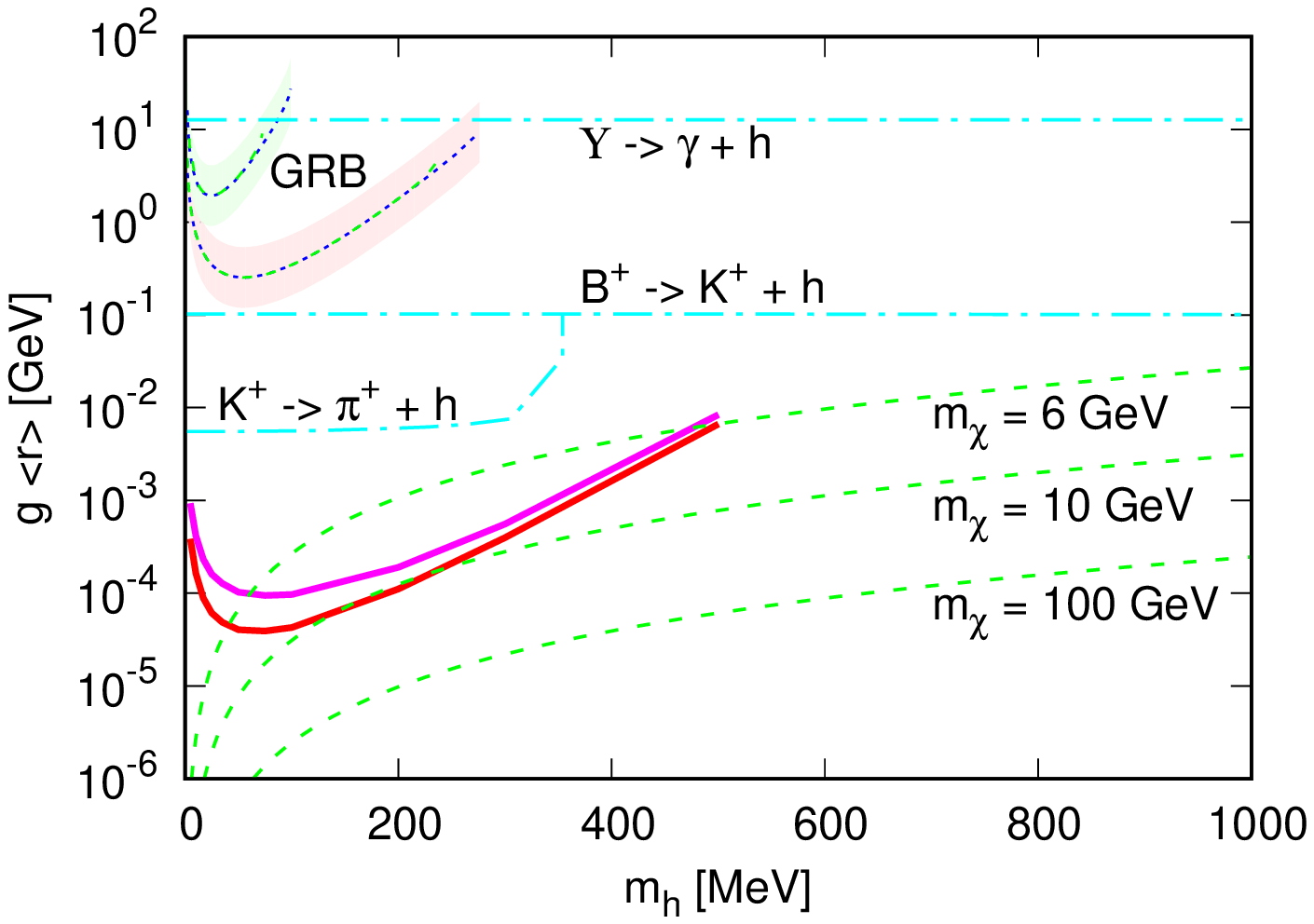}
\caption{Same as Fig.~\ref{fig:SNGRBmeson}, here including
the upper limits set by the dark matter direct search experiment LUX, 
for WIMP mass $M_\chi = 6$, $10$ and $100~\Gev$ (dashed lines, from top to bottom).}
\label{fig:DMSNGRBmeson}
\end{figure}
\end{center}

In the extended version of Weinberg's Higgs portal model, the spin-independent
WIMP-nucleon elastic scattering cross section is
(following the definition given in e.g. Ref.~\cite{Kurylov:2003ra}) 
\begin{equation}
   \sigma^{\rm SI}_{\chi N} =  \frac{4}{\pi}\, \left( \frac{1}{\sqrt{2}} \right)^2
   \mu^2_{\chi N} \left(\frac{f_\chi g \vevr f_N m_N}{m^2_H m^2_h} \right)^2\, .
\end{equation}
Here $\mu_{\chi N} = M_\chi m_N / (M_\chi + m_N)$ is the WIMP-nucleon reduced mass.
Latest exclusion limits published by the dark matter direct search experiments
LUX~\cite{Akerib:2016vxi}, PANDA-X~\cite{Tan:2016zwf}, and 
XENON1T~\cite{Aprile:2017iyp} are translated into
constraints on the parameter combination $f_\chi\, g\, \vevr / m^2_h$ for 
WIMP mass $M_\chi$ ranging from $6~\Gev$ to $1~\Tev$.
In order to make a comparison to the SN 1987A and laboratory constraints, 
the WIMP coupling is fixed by requiring the relic density to be 
$\Omega_\chi h^2 \simeq 0.11$, which yields 
$f_\chi \approx 0.02\, \sqrt{M_\chi}$~\cite{Anchordoqui:2013pta}.
The DM constraint was first derived in Ref.~\cite{Anchordoqui:2013bfa}, and here 
in Fig.~\ref{fig:DMSNGRBmeson} is shown for some representative values of WIMP mass 
$M_\chi = 6$, $10$ and $100~\Gev$.
Note that it does not become more stringent for larger WIMP masses, 
because the experimental limits on $\sigma^{\rm SI}_{\chi N}$ also scales 
approximately with $\sqrt{M_\chi}$ for $M_\chi \geq 100~\Gev$. 
We conclude that SN 1987A constraints are comparable to bounds from DM direct 
search results for $M_\chi \lesssim 10~\Gev$, while DM bounds for 
$M_\chi \gtrsim 100~\Gev$ are the strongest bounds among all on Weinberg's Higgs 
portal model.

\section{Summary}
\label{sec:summary}

Weinberg's Higgs portal model is another example to elucidate
that high-energy astrophysical objects such as the supernovae and gamma-ray bursts 
are excellent laboratory for particle physics.
In this model, massless Goldstone bosons arising from the spontaneous breaking of a $U(1)$ symmetry play the role of the dark radiation. 
The model was also extended to include a Majorana fermion of mass in the GeV to TeV 
range as the dark matter candidate.
Both particle species couple to the Standard Model fields solely through the SM 
Higgs boson.

Goldstone boson production in the proto-neutron star core is dominated by the 
emission of a real light Higgs boson in nuclear bremsstrahlung processes and its subsequent decay.
The SN 1987A constraint on Weinberg's Higgs portal model is obtained by finding the 
parameter regions for the Higgs portal coupling $g$, and the vacuum expectation value
of the light Higgs boson $\vevr$, for each light Higgs boson mass $m_h$, such that
the energy loss rate due to Goldstone boson emission satisfy the Raffelt criterion. 
In order to invoke this criterion, the Higgs portal coupling $g$ is required to be
smaller than the current collider bound inferred from the SM Higgs invisible decay, 
so that the Goldstone bosons are not trapped inside the proto-neutron star core.

We found that using the one-pion exchange (OPE) approximation and the SP07 global fits for the $pp$ elastic cross section results only in a factor of
$2.6$ difference for $10~\Mev \lesssim m_h \lesssim 50~\Mev$, and a factor of $1.4$
for $m_h > 300~\Mev$.
The SN 1987A constraints surpass those set by laboratory experiments or by energy 
loss arguments in other astrophysical objects, even with nuclear uncertainties taken 
into account.
In the extended version of Weinberg's Higgs portal model, latest exclusion limits published by the dark matter direct search experiments LUX, PANDA-X, and 
XENON1T are translated into constraints on the parameter combination $f_\chi\, g\, \vevr / m^2_h$ for WIMP mass $M_\chi$ ranging from $6~\Gev$ to $1~\Tev$.
Fixing the WIMP coupling $f_\chi$ with the measured dark matter relic density,
we found that SN 1987A constraints are comparable to bounds from DM direct 
search results for WIMP mass $M_\chi \lesssim 10~\Gev$, while DM bounds for 
$M_\chi \gtrsim 100~\Gev$ are the strongest bounds among all.

\section*{Acknowledgements}

We thank Xian-Wei Kang, Meng-Ru Wu, Tobias Fischer, and Jusak Tandean for the 
helpful discussions.
This work was supported in part by the Ministry of Science and Technology,
Taiwan, ROC under the Grant No. 104-2112-M-001-039-MY3.


\begin{thebibliography}{99}


\bibitem{Raffelt:1996wa}
  G.~G.~Raffelt,
  Chicago, USA: Univ. Pr. (1996) 664 p


\bibitem{Fischer:2009af}
  T.~Fischer, S.~C.~Whitehouse, A.~Mezzacappa, F.-K.~Thielemann and M.~Liebendorfer,
  Astron.\ Astrophys.\  {\bf 517} (2010) A80
  doi:10.1051/0004-6361/200913106
  [arXiv:0908.1871 [astro-ph.HE]].



\bibitem{Mueller:2014rna}
  B.~Müller and H.~T.~Janka,
  Astrophys.\ J.\  {\bf 788} (2014) 82
  doi:10.1088/0004-637X/788/1/82
  [arXiv:1402.3415 [astro-ph.SR]].


\bibitem{Camelio:2017nka}
  G.~Camelio, A.~Lovato, L.~Gualtieri, O.~Benhar, J.~A.~Pons and V.~Ferrari,
  arXiv:1704.01923 [astro-ph.HE].


\bibitem{Prakash:1996xs}
  M.~Prakash, I.~Bombaci, M.~Prakash, P.~J.~Ellis, J.~M.~Lattimer and R.~Knorren,
  Phys.\ Rept.\  {\bf 280} (1997) 1
  doi:10.1016/S0370-1573(96)00023-3
  [nucl-th/9603042].


\bibitem{Pons:1998mm}
  J.~A.~Pons, S.~Reddy, M.~Prakash, J.~M.~Lattimer and J.~A.~Miralles,
  Astrophys.\ J.\  {\bf 513} (1999) 780
  doi:10.1086/306889
  [astro-ph/9807040].


\cite{Nicotra:2005fj}
\bibitem{Nicotra:2005fj}
  O.~E.~Nicotra, M.~Baldo, G.~F.~Burgio and H.-J.~Schulze,
  Astron.\ Astrophys.\  {\bf 451} (2006) 213
  doi:10.1051/0004-6361:20053670
  [nucl-th/0506066].


\bibitem{Janka:2017vlw}
  H.-T.~Janka,
  arXiv:1702.08713 [astro-ph.HE].


\bibitem{Raffelt:1987yt}
  G.~Raffelt and D.~Seckel,
  Phys.\ Rev.\ Lett.\  {\bf 60} (1988) 1793.
  doi:10.1103/PhysRevLett.60.1793


\bibitem{Turner:1987by}
  M.~S.~Turner,
  Phys.\ Rev.\ Lett.\  {\bf 60} (1988) 1797.
  doi:10.1103/PhysRevLett.60.1797


\bibitem{Mayle:1987as}
  R.~Mayle, J.~R.~Wilson, J.~R.~Ellis, K.~A.~Olive, D.~N.~Schramm and G.~Steigman,
  Phys.\ Lett.\ B {\bf 203} (1988) 188.
  doi:10.1016/0370-2693(88)91595-X


\bibitem{Brinkmann:1988vi}
  R.~P.~Brinkmann and M.~S.~Turner,
  Phys.\ Rev.\ D {\bf 38} (1988) 2338.
  doi:10.1103/PhysRevD.38.2338


\bibitem{Janka:1995ir}
  H.~T.~Janka, W.~Keil, G.~Raffelt and D.~Seckel,
  Phys.\ Rev.\ Lett.\  {\bf 76} (1996) 2621
  doi:10.1103/PhysRevLett.76.2621
  [astro-ph/9507023].


\bibitem{Hanhart:2000er}
  C.~Hanhart, D.~R.~Phillips, S.~Reddy and M.~J.~Savage,
  Nucl.\ Phys.\ B {\bf 595} (2001) 335
  doi:10.1016/S0550-3213(00)00667-2
  [nucl-th/0007016].


\bibitem{Hanhart:2001fx}
  C.~Hanhart, J.~A.~Pons, D.~R.~Phillips and S.~Reddy,
  Phys.\ Lett.\ B {\bf 509} (2001) 1
  doi:10.1016/S0370-2693(01)00544-5
  [astro-ph/0102063].


\bibitem{Hannestad:2003yd}
  S.~Hannestad and G.~G.~Raffelt,
  Phys.\ Rev.\ D {\bf 67} (2003) 125008
   Erratum: [Phys.\ Rev.\ D {\bf 69} (2004) 029901]
  doi:10.1103/PhysRevD.69.029901, 10.1103/PhysRevD.67.125008
  [hep-ph/0304029].


\bibitem{Hannestad:2007ys}
  S.~Hannestad, G.~Raffelt and Y.~Y.~Y.~Wong,
  Phys.\ Rev.\ D {\bf 76} (2007) 121701
  doi:10.1103/PhysRevD.76.121701
  [arXiv:0708.1404 [hep-ph]].


\bibitem{Freitas:2007ip}
  A.~Freitas and D.~Wyler,
  JHEP {\bf 0712} (2007) 033
  doi:10.1088/1126-6708/2007/12/033
  [arXiv:0708.4339 [hep-ph]].


\bibitem{Chang:2016ntp}
  J.~H.~Chang, R.~Essig and S.~D.~McDermott,
  JHEP {\bf 1701} (2017) 107
  doi:10.1007/JHEP01(2017)107
  [arXiv:1611.03864 [hep-ph]].


\bibitem{Guha:2015kka}
  A.~Guha, Selvaganapathy J. and P.~K.~Das,
  Phys.\ Rev.\ D {\bf 95} (2017) no.1,  015001
  doi:10.1103/PhysRevD.95.015001
  [arXiv:1509.05901 [hep-ph]].


\bibitem{Ishizuka:1989ts}
  N.~Ishizuka and M.~Yoshimura,
  Prog.\ Theor.\ Phys.\  {\bf 84} (1990) 233.
  doi:10.1143/PTP.84.233


\bibitem{Arndt:2002yg}
  D.~Arndt and P.~J.~Fox,
  JHEP {\bf 0302} (2003) 036
  doi:10.1088/1126-6708/2003/02/036
  [hep-ph/0207098].


\bibitem{Keil:1996ju}
  W.~Keil, H.~T.~Janka, D.~N.~Schramm, G.~Sigl, M.~S.~Turner and J.~R.~Ellis,
  Phys.\ Rev.\ D {\bf 56} (1997) 2419
  doi:10.1103/PhysRevD.56.2419
  [astro-ph/9612222].
  

\bibitem{Fischer:2016cyd}
  T.~Fischer, S.~Chakraborty, M.~Giannotti, A.~Mirizzi, A.~Payez and A.~Ringwald,
  arXiv:1605.08780 [astro-ph.HE].


\bibitem{Raffelt:1990yz}
  G.~G.~Raffelt,
  Phys.\ Rept.\  {\bf 198}, 1 (1990).
  

\bibitem{Raffelt:2006cw}
  G.~G.~Raffelt,
  Lect.\ Notes Phys.\  {\bf 741}, 51 (2008)
  [hep-ph/0611350].


\bibitem{Weinberg:2013kea}
  S.~Weinberg,
  Phys.\ Rev.\ Lett.\  {\bf 110} (2013) no.24,  241301
  doi:10.1103/PhysRevLett.110.241301
  [arXiv:1305.1971 [astro-ph.CO]].


\bibitem{Riess:2016jrr}
  A.~G.~Riess {\it et al.},
  Astrophys.\ J.\  {\bf 826} (2016) no.1,  56
  doi:10.3847/0004-637X/826/1/56
  [arXiv:1604.01424 [astro-ph.CO]].
  
  
\bibitem{Heavens:2017hkr}
  A.~Heavens, Y.~Fantaye, E.~Sellentin, H.~Eggers, Z.~Hosenie, S.~Kroon and A.~Mootoovaloo,
  arXiv:1704.03467 [astro-ph.CO].


\bibitem{Ng:2014iqa}
  K.~W.~Ng, H.~Tu and T.~C.~Yuan,
  JCAP {\bf 1409} (2014) no.09,  035
  doi:10.1088/1475-7516/2014/09/035
  [arXiv:1406.1993 [hep-ph]].


\bibitem{Cheung:2013oya}
  K.~Cheung, W.~Y.~Keung and T.~C.~Yuan,
  Phys.\ Rev.\ D {\bf 89} (2014) no.1,  015007
  doi:10.1103/PhysRevD.89.015007
  [arXiv:1308.4235 [hep-ph]].


\bibitem{Anchordoqui:2013bfa}
  L.~A.~Anchordoqui, P.~B.~Denton, H.~Goldberg, T.~C.~Paul, L.~H.~M.~Da Silva, B.~J.~Vlcek and T.~J.~Weiler,
  Phys.\ Rev.\ D {\bf 89} (2014) no.8,  083513
  doi:10.1103/PhysRevD.89.083513
  [arXiv:1312.2547 [hep-ph]].


\bibitem{Akerib:2016vxi}
  D.~S.~Akerib {\it et al.},
  arXiv:1608.07648 [astro-ph.CO].


\bibitem{Aprile:2017iyp}
  E.~Aprile {\it et al.} [XENON Collaboration],
  arXiv:1705.06655 [astro-ph.CO].


\bibitem{Keung:2013mfa}
  W.~Y.~Keung, K.~W.~Ng, H.~Tu and T.~C.~Yuan,
  Phys.\ Rev.\ D {\bf 90} (2014) no.7,  075014
  doi:10.1103/PhysRevD.90.075014
  [arXiv:1312.3488 [hep-ph]].


\bibitem{Tu:2015lwv}
  H.~Tu and K.~W.~Ng,
  JCAP {\bf 1603} (2016) no.03,  037
  doi:10.1088/1475-7516/2016/03/037
  [arXiv:1512.05165 [hep-ph]].


\bibitem{Drees:1993bu}
  M.~Drees and M.~Nojiri,
  Phys.\ Rev.\ D {\bf 48} (1993) 3483
  doi:10.1103/PhysRevD.48.3483
  [hep-ph/9307208].


\bibitem{Jungman:1995df}
  G.~Jungman, M.~Kamionkowski and K.~Griest,
  Phys.\ Rept.\  {\bf 267} (1996) 195
  doi:10.1016/0370-1573(95)00058-5
  [hep-ph/9506380].


\bibitem{Hisano:2011cs}
  J.~Hisano, K.~Ishiwata, N.~Nagata and T.~Takesako,
  JHEP {\bf 1107} (2011) 005
  doi:10.1007/JHEP07(2011)005
  [arXiv:1104.0228 [hep-ph]].


\bibitem{Cheng:2012qr}
  H.~Y.~Cheng and C.~W.~Chiang,
  JHEP {\bf 1207} (2012) 009
  doi:10.1007/JHEP07(2012)009
  [arXiv:1202.1292 [hep-ph]].


\bibitem{Cline:2013gha}
  J.~M.~Cline, K.~Kainulainen, P.~Scott and C.~Weniger,
  Phys.\ Rev.\ D {\bf 88} (2013) 055025
   Erratum: [Phys.\ Rev.\ D {\bf 92} (2015) no.3,  039906]
  doi:10.1103/PhysRevD.92.039906, 10.1103/PhysRevD.88.055025
  [arXiv:1306.4710 [hep-ph]].



\bibitem{He:2013suk}
  X.~G.~He and J.~Tandean,
  Phys.\ Rev.\ D {\bf 88} (2013) 013020
  doi:10.1103/PhysRevD.88.013020
  [arXiv:1304.6058 [hep-ph]].


\bibitem{Anchordoqui:2013pta}
  L.~A.~Anchordoqui and B.~J.~Vlcek,
  Phys.\ Rev.\ D {\bf 88} (2013) 043513
  doi:10.1103/PhysRevD.88.043513
  [arXiv:1305.4625 [hep-ph]].


\bibitem{Bechtle:2014ewa}
  P.~Bechtle, S.~Heinemeyer, O.~Stål, T.~Stefaniak and G.~Weiglein,
  JHEP {\bf 1411} (2014) 039
  doi:10.1007/JHEP11(2014)039
  [arXiv:1403.1582 [hep-ph]].


\bibitem{Taketani:1951}
M.~Taketani, S.~Nakamura and M.~Sasaki, 
Prog. Theor. Phys. 6 (1951), 581.


\bibitem{Entem:2017gor}
  D.~R.~Entem, R.~Machleidt and Y.~Nosyk,
  arXiv:1703.05454 [nucl-th].


\bibitem{Entem:2015xwa}
  D.~R.~Entem, N.~Kaiser, R.~Machleidt and Y.~Nosyk,
  Phys.\ Rev.\ C {\bf 92} (2015) no.6,  064001
  doi:10.1103/PhysRevC.92.064001
  [arXiv:1505.03562 [nucl-th]].


\bibitem{Machleidt:2001rw}
  R.~Machleidt and I.~Slaus,
  J.\ Phys.\ G {\bf 27} (2001) R69
  doi:10.1088/0954-3899/27/5/201
  [nucl-th/0101056].


\bibitem{Naghdi:2007ek}
  M.~Naghdi,
  Phys.\ Part.\ Nucl.\  {\bf 45} (2014) 924
  doi:10.1134/S1063779614050050
  [nucl-th/0702078].


\bibitem{Machleidt:2011zz}
  R.~Machleidt and D.~R.~Entem,
  Phys.\ Rept.\  {\bf 503} (2011) 1
  doi:10.1016/j.physrep.2011.02.001
  [arXiv:1105.2919 [nucl-th]].


\bibitem{Machleidt:2016vlh}
  R.~Machleidt,
  Symmetry {\bf 8} (2016) no.4,  26.
  doi:10.3390/sym8040026


\bibitem{Bacca:2008yr}
  S.~Bacca, K.~Hally, C.~J.~Pethick and A.~Schwenk,
  Phys.\ Rev.\ C {\bf 80} (2009) 032802
  doi:10.1103/PhysRevC.80.032802
  [arXiv:0812.0102 [nucl-th]].


\bibitem{Bacca:2015tva}
  S.~Bacca, R.~Sharma and A.~Schwenk,
  arXiv:1509.08151 [nucl-th].


\bibitem{Friman:1978zq}
  B.~L.~Friman and O.~V.~Maxwell,
  Astrophys.\ J.\  {\bf 232} (1979) 541.
  doi:10.1086/157313


\bibitem{Hannestad:1997gc}
  S.~Hannestad and G.~Raffelt,
  Astrophys.\ J.\  {\bf 507} (1998) 339
  doi:10.1086/306303
  [astro-ph/9711132].


\bibitem{Bartl:2016iok}
  A.~Bartl, R.~Bollig, H.~T.~Janka and A.~Schwenk,
  Phys.\ Rev.\ D {\bf 94} (2016) 083009
  doi:10.1103/PhysRevD.94.083009
  [arXiv:1608.05037 [nucl-th]].


\bibitem{Bartl:2014hoa}
  A.~Bartl, C.~J.~Pethick and A.~Schwenk,
  Phys.\ Rev.\ Lett.\  {\bf 113} (2014) 081101
  doi:10.1103/PhysRevLett.113.081101
  [arXiv:1403.4114 [nucl-th]].


\bibitem{Hanhart:2000ae}
  C.~Hanhart, D.~R.~Phillips and S.~Reddy,
  Phys.\ Lett.\ B {\bf 499} (2001) 9
  doi:10.1016/S0370-2693(00)01382-4
  [astro-ph/0003445].


\bibitem{Limkaisang:2001yz}
  V.~Limkaisang, K.~Harada, J.~Nagata, H.~Yoshino, Y.~Yoshino, M.~Shoji and M.~Matsuda,
  Prog.\ Theor.\ Phys.\  {\bf 105} (2001) 233.
  doi:10.1143/PTP.105.233


\bibitem{Babenko:2016idp}
  V.~A.~Babenko and N.~M.~Petrov,
  arXiv:1604.02912 [nucl-th].


\bibitem{MartinezPinedo:2012rb}
  G.~Martinez-Pinedo, T.~Fischer, A.~Lohs and L.~Huther,
  Phys.\ Rev.\ Lett.\  {\bf 109} (2012) 251104
  doi:10.1103/PhysRevLett.109.251104
  [arXiv:1205.2793 [astro-ph.HE]].


\bibitem{Baldo:2016jhp}
  M.~Baldo and G.~F.~Burgio,
  arXiv:1606.08838 [nucl-th].


\bibitem{Trautmann:2016ntm}
  W.~Trautmann, M.~D.~Cozma and P.~Russotto,
  PoS Bormio {\bf 2016} (2016) 036
  [arXiv:1610.03650 [nucl-ex]].


\bibitem{Bacca:2011qd}
  S.~Bacca, K.~Hally, M.~Liebendorfer, A.~Perego, C.~J.~Pethick and A.~Schwenk,
  Astrophys.\ J.\  {\bf 758} (2012) 34
  doi:10.1088/0004-637X/758/1/34
  [arXiv:1112.5185 [astro-ph.HE]].


\bibitem{Albers:2004iw}
  D.~Albers {\it et al.},
  Eur.\ Phys.\ J.\ A {\bf 22} (2004) 125
  doi:10.1140/epja/i2004-10011-3
  [nucl-ex/0403045].


\bibitem{Wilkin:2016qio}
  C.~Wilkin,
  EPJ Web Conf.\  {\bf 130} (2016) 01007.
  doi:10.1051/epjconf/201613001007


\bibitem{Arndt:2000xc}
  R.~A.~Arndt, I.~I.~Strakovsky and R.~L.~Workman,
  Phys.\ Rev.\ C {\bf 62} (2000) 034005
  doi:10.1103/PhysRevC.62.034005
  [nucl-th/0004039].


\bibitem{Nijmegen}
http://nn-online.org


\bibitem{SAID}
http://gwdac.phys.gwu.edu


\bibitem{Kang:2014ioa}
  X.~W.~Kang, PhD thesis, 
 ``Chiral Dynamics and Final State Interactions in Semileptonic B Meson Decay and Antinucleon-Nucleon Scattering,'' University of Bonn (2014),
 http://hss.ulb.uni-bonn.de/2014/3714/3714.htm


\bibitem{Arndt:2007qn}
  R.~A.~Arndt, W.~J.~Briscoe, I.~I.~Strakovsky and R.~L.~Workman,
  Phys.\ Rev.\ C {\bf 76} (2007) 025209
  doi:10.1103/PhysRevC.76.025209
  [arXiv:0706.2195 [nucl-th]].


\bibitem{Arndt:2008uc}
  R.~A.~Arndt, W.~J.~Briscoe, A.~B.~Laptev, I.~I.~Strakovsky and R.~L.~Workman,
  Nucl.\ Sci.\ Eng.\  {\bf 162} (2009) 312
  [arXiv:0806.1198 [nucl-ex]].


\bibitem{Konobeevski:2017mpw}
  E.~S.~Konobeevski, S.~V.~Zuyev, V.~I.~Kukulin and V.~N.~Pomerantsev,
  arXiv:1703.00519 [nucl-th].



\bibitem{Stoks:1993tb}
  V.~G.~J.~Stoks, R.~A.~M.~Klomp, M.~C.~M.~Rentmeester and J.~J.~de Swart,
  Phys.\ Rev.\ C {\bf 48} (1993) 792.
  doi:10.1103/PhysRevC.48.792


\bibitem{Low:1958sn}
  F.~E.~Low,
  Phys.\ Rev.\  {\bf 110} (1958) 974.
  doi:10.1103/PhysRev.110.974


\bibitem{Adler:1966gc}
  S.~L.~Adler and Y.~Dothan,
  Phys.\ Rev.\  {\bf 151} (1966) 1267.
  doi:10.1103/PhysRev.151.1267


\bibitem{Heller:1969ur}
  L.~Heller,
  Phys.\ Rev.\  {\bf 174} (1968) 1580.
  doi:10.1103/PhysRev.174.1580



\bibitem{Burrows:1990pk}
  A.~Burrows, M.~T.~Ressell and M.~S.~Turner,
  Phys.\ Rev.\ D {\bf 42} (1990) 3297.
  doi:10.1103/PhysRevD.42.3297


\bibitem{Tubbs:1975jx}
  D.~L.~Tubbs and D.~N.~Schramm,
  Astrophys.\ J.\  {\bf 201} (1975) 467.
  doi:10.1086/153909


\bibitem{Guillot:2013wu}
  S.~Guillot, M.~Servillat, N.~A.~Webb and R.~E.~Rutledge,
  Astrophys.\ J.\  {\bf 772} (2013) 7
  doi:10.1088/0004-637X/772/1/7
  [arXiv:1302.0023 [astro-ph.HE]].


\bibitem{Raithel:2016vtt}
  C.~A.~Raithel, F.~Ozel and D.~Psaltis,
  Phys.\ Rev.\ C {\bf 93} (2016) no.3,  032801
  doi:10.1103/PhysRevC.93.032801
  [arXiv:1603.06594 [astro-ph.HE]].


\bibitem{Lattimer:2015nhk}
  J.~M.~Lattimer and M.~Prakash,
  arXiv:1512.07820 [astro-ph.SR].


\bibitem{Miller:2016pom}
  M.~C.~Miller and F.~K.~Lamb,
  Eur.\ Phys.\ J.\ A {\bf 2016} 52
  [arXiv:1604.03894 [astro-ph.HE]].


\bibitem{Baldo:2014yja}
  M.~Baldo, G.~F.~Burgio, H.-J.~Schulze and G.~Taranto,
  Phys.\ Rev.\ C {\bf 89} (2014) no.4,  048801
  doi:10.1103/PhysRevC.89.048801
  [arXiv:1404.7031 [nucl-th]].



\bibitem{Bird:2004ts}
  C.~Bird, P.~Jackson, R.~V.~Kowalewski and M.~Pospelov,
  Phys.\ Rev.\ Lett.\  {\bf 93} (2004) 201803
  doi:10.1103/PhysRevLett.93.201803
  [hep-ph/0401195].


\bibitem{Huang:2013oua}
  F.~P.~Huang, C.~S.~Li, D.~Y.~Shao and J.~Wang,
  Eur.\ Phys.\ J.\ C {\bf 74} (2014) 8,  2990
  [arXiv:1307.7458 [hep-ph]].


\bibitem{delAmoSanchez:2010bk}
  P.~del Amo Sanchez {\it et al.} [BaBar Collaboration],
  Phys.\ Rev.\ D {\bf 82} (2010) 112002
  doi:10.1103/PhysRevD.82.112002
  [arXiv:1009.1529 [hep-ex]].



\bibitem{Artamonov:2009sz}
  A.~V.~Artamonov {\it et al.} [BNL-E949 Collaboration],
  Phys.\ Rev.\ D {\bf 79} (2009) 092004
  doi:10.1103/PhysRevD.79.092004
  [arXiv:0903.0030 [hep-ex]].


\bibitem{Aaij:2016qsm}
  R.~Aaij {\it et al.} [LHCb Collaboration],
  Phys.\ Rev.\ D {\bf 95} (2017) no.7,  071101
  doi:10.1103/PhysRevD.95.071101
  [arXiv:1612.07818 [hep-ex]].


\bibitem{Kurylov:2003ra}
  A.~Kurylov and M.~Kamionkowski,
  Phys.\ Rev.\ D {\bf 69} (2004) 063503
  doi:10.1103/PhysRevD.69.063503
  [hep-ph/0307185].


\bibitem{Tan:2016zwf}
  A.~Tan {\it et al.} [PandaX-II Collaboration],
  Phys.\ Rev.\ Lett.\  {\bf 117} (2016) no.12,  121303
  doi:10.1103/PhysRevLett.117.121303
  [arXiv:1607.07400 [hep-ex]].



\end{thebibliography}

\end{document}